\documentclass[conf]{IEEEtran}
\usepackage{algorithm,algpseudocode}
\usepackage{amsmath}
\usepackage{amssymb}
\usepackage{mathtools}
\usepackage[dvipsnames]{xcolor}

\usepackage{pifont}
\usepackage{cite}
\def\BibTeX{{\rm B\kern-.05em{\sc i\kern-.025em b}\kern-.08em
    T\kern-.1667em\lower.7ex\hbox{E}\kern-.125emX}}
\usepackage[bookmarksnumbered=true]{hyperref} 
\hypersetup{
     colorlinks = true,
     linkcolor = blue,
     anchorcolor = blue,
     citecolor = blue,
     filecolor = blue,
     urlcolor = blue
     }

\usepackage{graphicx}
\usepackage{svg}

\usepackage{wrapfig}
\usepackage{caption}
\usepackage{subcaption}
\usepackage{booktabs}
\newcommand*\circled[1]{\tikz[baseline=(char.base)]{
            \node[shape=circle,draw,inner sep=1pt] (char) {#1};}}
\usepackage{tikz}

\usepackage{ragged2e}
\usepackage{siunitx}
\usepackage{multicol}
\usepackage{multirow}

\usepackage{xcolor}
\begin{document}

\title{RobustEdge: Low Power Adversarial Detection for Cloud-Edge Systems}

\author{ Abhishek Moitra, \textit{Student Member, IEEE}, Abhiroop Bhattacharjee, \textit{Student Member, IEEE}, Youngeun Kim, \textit{Student Member, IEEE} and Priyadarshini Panda, \textit{Member, IEEE}
\thanks{Abhishek Moitra, Abhiroop Bhattacharjee, Youngeun Kim and Priyadarshini Panda are with
the Department of Electrical Engineering, Yale University, New Haven, CT,
USA. }
}

\maketitle

\begin{abstract}
  In practical cloud-edge scenarios, where a resource constrained edge performs data acquisition and a cloud system (having sufficient resources) performs inference tasks with a deep neural network (DNN), adversarial robustness is critical for reliability and ubiquitous deployment. Adversarial detection is a prime adversarial defense technique used in prior literature. However, in prior detection works, the detector is attached to the classifier model and both detector and classifier work in tandem to perform adversarial detection that requires a high computational overhead which is not available at the low-power edge. Therefore, prior works can only perform adversarial detection at the cloud and not at the edge. This means that in case of adversarial attacks, the unfavourable adversarial samples must be communicated to the cloud which leads to energy wastage at the edge device. Therefore, a low-power edge-friendly adversarial detection method is required to improve the energy efficiency of the edge and robustness of the cloud-based classifier. To this end, RobustEdge proposes \textit{Quantization-enabled Energy Separation} (QES) training with ``early detection and exit" to perform edge-based low cost adversarial detection. The QES-trained detector implemented at the edge blocks adversarial data transmission to the classifier model, thereby improving adversarial robustness and energy-efficiency of the Cloud-Edge system. Through extensive experiments on CIFAR10, CIFAR100 and TinyImagenet, we find that 16-bit and 12-bit quantized detectors achieve a high AUC score $>$ 0.9 across different datasets and adversarial attacks. Hardware evaluations on a 45nm CMOS digital accelerator reveal that RobustEdge is requires 25$\times$ lower energy for adversarial detection compared to prior works. Additionally, compared to prior works that perform adversarial detection at the cloud, we find that edge-based adversarial detection can improve the energy-efficiency of the cloud-edge system by $>166\times$. Furthermore, we find that RobustEdge is transferable across datasets \textit{i.e.,} a detector trained on one dataset can detect adversaries on another dataset.
\end{abstract}



\begin{IEEEkeywords}Adversarial Attacks, Adversarial Detection, Edge Computing, Energy Efficiency 
\end{IEEEkeywords}

\maketitle

\section{Introduction}
{In the era of IoT and edge computing, deep neural networks (DNNs) are being ubiquitously implemented in cloud-edge systems \cite{madria2013sensor, eshratifar2019bottlenet, othman2013survey, xue2021ddpqn, xue2021eosdnn}. Fig. \ref{fig:motivation}a shows an example of a cloud-edge system where a low powered edge device performs data acquisition and transmits the acquired data to a classifier/DNN model deployed at the cloud for inference \cite{yu2017computation, huang2018distributed}. However, recent works have found that DNNs are vulnerable to adversarial attacks, wherein adding small structured noise to the input data can fool the classifier model \cite{huang2017adversarial, madry2017towards}. In the cloud edge scenario shown in Fig. \ref{fig:motivation}a, the cloud-based classifier model is prone to such adversarial attacks. To this end, several adversarial detection methods \cite{metzen2017detecting, sterneck2021noise, xu2017feature, moitra2021detectx, grosse2017statistical} have been researched to avert adversarial attacks. 

However, prior detection works have broadly followed two different approaches- (Case 1) Works such as \cite{grosse2017statistical, moitra2021detectx} train the classifier network on both adversarial and natural samples in order to perform adversarial detection; (Case 2) Another approach is to train small detector networks on intermediate activations of classifier models \cite{metzen2017detecting, sterneck2021noise, xu2017feature} to distinguish natural or adversarial samples. In both cases, the adversarial detection always occurs at the classifier model. This has the following connotations: 1) it leads to huge size of classifier-detector models and 2) they entail high number of computations leading to high energy overhead. These factors render prior detection approaches unsuitable for resource-constrained edge deployment. Therefore, prior works are cloud-centric adversarial detection solutions. As seen in Fig. \ref{fig:motivation}b, cloud-centric adversarial detection entails wasteful energy expenditure by the edge device to transmit adversarial data to the cloud. Therefore, an edge-friendly adversarial detection method that requires extremely small detector model size and computations for adversarial detection is critical for improving the adversarial robustness. Additionally, adversarial detection at the edge will eliminate wasteful transmission energy consumption in cloud-edge systems.

\begin{figure}[t]
    \centering
    \begin{tabular}{c}
         \includegraphics[width=\columnwidth]{ 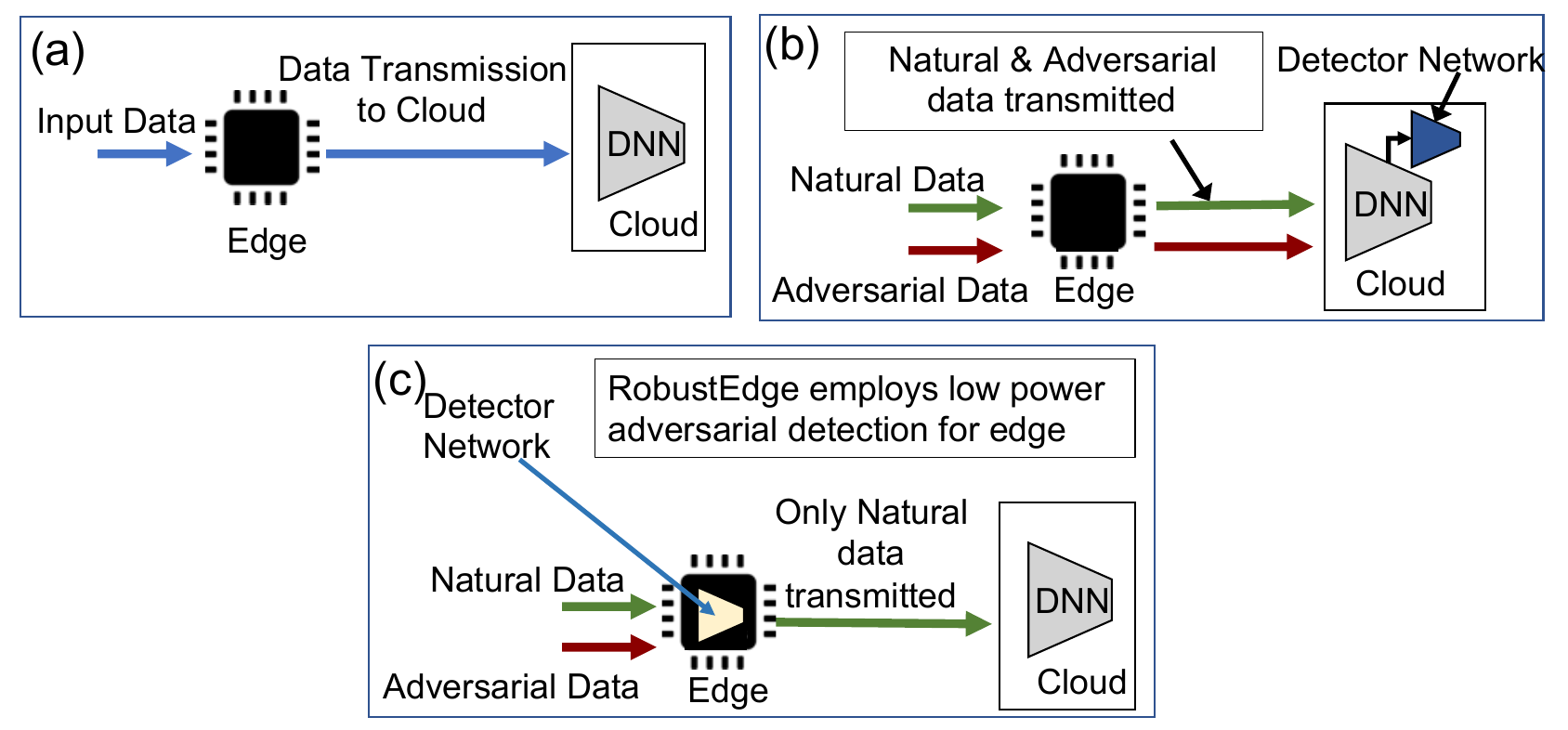}
    \end{tabular}
    
    \caption{(a) A standard Cloud-Edge system where the edge performs data acquisition and the classifier DNN on cloud performs inference. (b) Prior classifier-based detection methods are not suitable for edge deployment and entail higher edge to cloud data transmission causing wasteful energy expenditure at the edge. (c) RobustEdge trains a small energy-efficient detector suitable for low power edge deployment. The detector minimizes adversarial data and maximizes natural data transmission to cloud.}
    \label{fig:motivation}
\end{figure}

To this end, we propose RobustEdge that uses \textit{quantization-enabled energy separation} (QES) training to train a small, standalone adversarial detector that can be suitably deployed at a low power edge platform as shown in Fig. \ref{fig:motivation}c. Note, here \textit{{energy separation}} does not mean the actual hardware energy but is a characterization function to distinguish adversarial and natural inputs. Additionally, we propose an ``early detection and exit" strategy to improve the compute efficiency of the adversarial detection. The detector successfully detects adversarial and natural samples at the edge. The natural samples are transmitted to the cloud while the adversarial data transmission is terminated at the edge minimizing the wasteful energy expenditure.  

In summary the key contributions of our paper are as follows:
\begin{enumerate}
    \item We propose RobustEdge which is the first of its kind low power edge-based adversarial detection technique that employs a novel QES training method with ``early detection and exit" strategy to achieve high performance and compute efficient adversarial detection.
    \item Based on extensive evaluations using benchmark datasets- CIFAR10, CIFAR100, TinyImagenet and comprehensive adversarial attacks, we find that 16-bit and 12-bit detectors achieve high detection performance across different gradient-based (AUC score $>$ 0.9) and score-based attacks (AUC score $>$ 0.7). Additionally, the 16-bit and 12-bit detectors eliminate 100\% of the edge-cloud data transmission for adversarial inputs which minimizes the energy wastage at the edge by $\sim70$J.
    
    \item We implement the QES-trained detector on a 45nm digital custom CMOS hardware accelerator. We find that RobustEdge requires $<25\times$ energy for adversarial detection compared to prior works. Furthermore, we find that ``early detection and exit" strategy leads to 66\% lower detection energy compared to detection without early-exit. 
    \item Interestingly, we find that the QES-trained detector is transferable across datasets. A detector trained on the TinyImagenet dataset can detect different adversarial attacks on the CIFAR10 and CIFAR100 dataset (AUC $>$ 0.9 and AUC $>$ 0.7 for gradient and score-based attacks, respectively).
\end{enumerate}
Note, in this paper, we focus on practical cloud-edge systems (where, the resource constrained edge is only responsible for data collection and transmission) that find use in real applications today, e.g., mobile phones, voice assistants, autonomous cars and drones etc \cite{zheng2018dynamic, sundararaj2019optimal}. {There have been many recent works \cite{li2018edge,hu2019dynamic,he2020joint,xiong2023multi} proposing cloud-edge systems, where, the edge is assumed to have sufficient compute resources to perform partial inference. These works use compute offloading to the edge system which send the data or intermediate activations to the cloud.} RobustEdge targets the former practical use case of cloud-edge system, and strengthens the edge with ultra low power and cost-friendly adversarial detection. 

\section{Background}

\subsection{Background on Adversarial Attacks}
\label{background_adv_attacks}
Adversarial attacks have the following objective: $$C(W, x+\delta) \neq C(W, x), ~||\delta|| \leq \epsilon.$$ 

Here, $C(.)$ denotes the classifer model's predictions, $W$ is the classifier weights, $x$ is the natural input data and $\delta$ is the perturbation added to the input. Based on the input, $\delta$ is computed using the gradient (in case of gradient-based attacks) and loss (decision-based attacks) information from the classifier model \cite{huang2017adversarial, madry2017towards, andriushchenko2020square}. $\delta$ is constrained by the parameter $\epsilon$ such that attacks are imperceptible to human eyes. 

There have been several effective adversarial attacks proposed in literature which are described below:
\begin{enumerate}

\item \textbf{Gradient-based attacks:} To generate these attacks, a forward and backward propagation is performed on the classifier model. To create the adversarial image for input $x$, the gradient of loss with respect to $x$ ($\frac{\partial \mathcal{L}}{\partial x}$) is used. The Fast Gradient Sign Method (FGSM) is a simple one-step adversarial attack proposed in \cite{kurakin2016adversarial}. Several works have shown that FGSM attack can be made stronger with momentum (MIFGSM)\cite{dong2018boosting}, random initialization (FFGSM)\cite{wong2020fast}, and input diversification (DIFGSM)\cite{xie2019improving}. In contrast, the Basic Iterative Method (BIM) is an iterative attack proposed in \cite{kurakin2016adversarial}. The BIM attack with random restarts is called the Projected Gradient Descent (PGD) attack \cite{madry2017towards}. A targeted version of the PGD attack (TPGD) \cite{zhang2019theoretically} can fool the model into mis-classifying a data as a desired class. Other multi-step attacks like Carlini-Wagner (C\&W) \cite{carlini2017towards} and PGD-L2 \cite{madry2017towards} are crafted by computing the L2 Norm distance between the adversarial and natural images.


\item \textbf{Score-based attacks:} Score-based attacks do not require input gradients to craft adversaries. Instead these attacks use the classifier's loss information to maximize the attack strength. Square attack (SQR) \cite{andriushchenko2020square} uses multiple queries to perturb randomly selected square regions in the input. Other score-based attacks like the AutoAttack (AUTO) and Auto-PGD (APGD) craft adversaries by automatically choosing the optimal attack parameters \cite{croce2020reliable}. The Gaussian Noise (GN) attack is created by adding gaussian noise with standard deviation of strenght $\epsilon$ to the input.


\end{enumerate}




\subsection{Eyeriss DNN inference accelerator}

The Eyeriss accelerator is a digital hardware architecture designed for energy-efficient DNN inference that performs high-speed convolution operations \cite{chen2016eyeriss}. The Eyeriss architecture as shown in Fig. \ref{fig:eyeriss_arch} is based on a systolic array design, which performs parallel computing using a grid of processing elements (PEs) to emulate Multiply-and-Accumulate (MAC) operations on hardware. Inside each PE, there are registers or scratchpads (spads) for storing input activations, weights \& partial sums, digital multipliers and accumulators for MAC computations, digital comparators for ReLU activation and a PE control unit, all of which are optimized for efficient matrix operations. Eyeriss uses a technique called `dataflow mapping' to distribute computation across the systolic array of PEs. These dataflows, essentially, partition the input activations and the convolution filters stored in the caches into smaller chunks that can be processed in parallel by the PEs with minimum communication overhead in the data transfer. With `dataflow mapping', Eyeriss can perform convolutions faster and more efficiently than traditional CPU or GPU architectures by minimizing data movement and maximizing parallelism. In this work, an output-stationary (OS) dataflow is used in carrying out MAC operations in the PE array that helps reduce repeated accesses of the intermediate partial sums to and from the main memory.

\begin{figure}[t]
    \centering
   
         \includegraphics[width=\columnwidth]{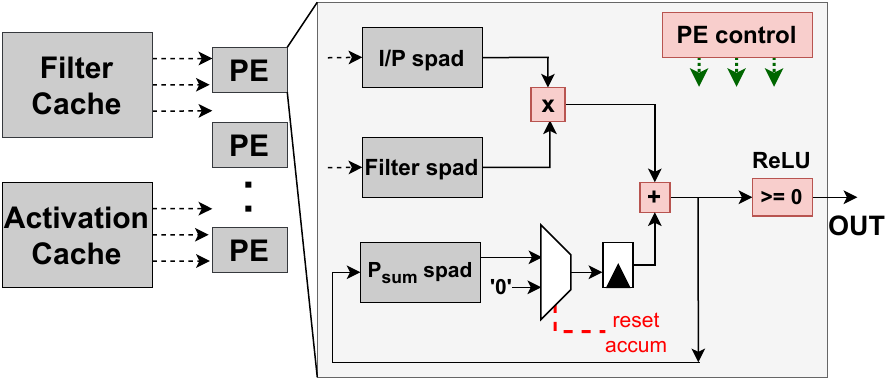}

    \caption{An overview of the Eyeriss architecture consisting of a systolic array of PEs for DNN inference.}
    \label{fig:eyeriss_arch}
\end{figure}

\section{Related Works}
Predominantly, there are two types of adversarial defense strategies used in prior literature. 1) Adversarial Classification 2) Adversarial Detection. 
\subsection{Adversarial Classification}
Here, works such as \cite{guo2020meets} proposed input feature transformation using JPEG compression followed by training on compressed feature space to improve the classification performance of the classifier model. Madry et al. \cite{madry2017towards} proposed adversarial training in which a classifier model is trained on adversarial and clean data to improve the adversarial and clean classification performance. Following this, several works have used noise injection into parameters \cite{he2019parametric} and ensemble adversarial training to harden the classifier model against a wide range of attacks. Lin et al. \cite{lin2019defensive} showed that adversarial classification can be improved by reducing the error amplification in a network. Hence, they used adversarial training with regularization to constrain the Lipschitz constant of the network to less than unity. However, for the kind of practical cloud-edge scenario considered in this work adversarial training methods are not suitable as the classifier model at the cloud requires modification for different attacks.
\subsection{Adversarial Detection}
\label{rel_works_training}
Towards adversarial detection, a line of works have trained the classifier model for adversarial detection.
Xu et al. \cite{xu2017feature} propose a method that uses outputs of multiple classifier models to estimate the difference between natural and adversarial data. Here, the classifier models are trained on natural inputs with different feature squeezing techniques at the inputs. Moitra et al. \cite{moitra2021detectx} uses the features from the first layer of the classifier model to perform adversarial detection. In particular, they perform adversarial detection using hardware signatures in analog crossbar accelerators. While Grosse et al. \cite{grosse2017statistical} train the classifier model with an additional class label indicating adversarial data, Gong et al. \cite{gong2017adversarial} train a separate binary classifier on the natural and adversarial data generated from the classifier model to perform adversarial detection. Lee et al. \cite{lee2018simple} use a metric called the \textit{Mahalanobis distance} classifier to train the classifier model. The Mahalanobis distance is used to distinguish natural from adversarial data.    

Other works have trained detector networks attached to classifier models for adversarial detection.
Metzen et al. \cite{metzen2017detecting} and Sterneck et al. \cite{sterneck2021noise} use the intermediate features of the classifier model to train a simple binary adversarial detector. While Metzen et al. \cite{metzen2017detecting} use a heuristic-based method to determine the point of attachment of the detector with the classifier model, Sterneck et al. \cite{sterneck2021noise} use a structured metric called adversarial noise sensitivity to do the same. Similarly, Yin et al. \cite{yin2019gat} use asymmetric adversarial training to train detectors on the intermediate features of the classifier model for adversarial detection. Further, Huang et al. \cite{huang2019model} use the confidence scores from the classifier model to estimate the \textit{relative score difference} corresponding to the clean and the adversarial input to perform adversarial detection. Further, they also recommend classifier model training on noisy data to improve the adversarial detection performance.   

Although prior detection works have achieved state-of-the-art performance, they have overlooked the practicality and the implication of such techniques for cloud-edge computing scenarios. As the detector is attached to the classifier, 1) Adversarial detection cost is high as both classifier and detector work in tandem. 2) The detector needs to be retrained if the classifier model changes. RobustEdge performs low-cost adversarial detection while being agnostic of the classifier DNN deployed at the cloud. At the same time, the detection occurs at the edge eliminating wasteful adversarial data transmission to the cloud. 

\section{Methodology}

\label{algorithm}
\begin{algorithm}
\hspace*{\algorithmicindent} \textbf{Input} $n$ layered detector ($\mathcal{D}$), $x_{nat}$, $x_{adv}$, $s_{nat}$, $K$, $L$, $U$\\
\hspace*{\algorithmicindent} \textbf{Output} Trained detector $\mathcal{D}_T$, $\mathcal{E}_{n,Th}$, $\mathcal{E}_{i,L}$, $\mathcal{E}_{i,U}$
\begin{algorithmic}[1]
\caption{\textit{QES Training Algorithm}}
\label{alg:algorithm}
    \State{\textcolor{violet}{/* Detector Weight Training */}}
    \ForAll {i = 1 to n}
        \State {Quantize layers $[1,i]$ to k-bit} 
        \State {Freeze layers $[1,i$-$1]$}
        \ForAll {j = 1 to $N_{epoch}$}
            \ForAll {B = 1 to $N_{batches}$}
            \State {$X_n^B$, $X_a^B$ $\gets$ Mini-batch of $x_{nat}$ and $x_{adv}$}
            \State {Compute $\mathcal{E}_{nat}^{i,B}$ and $\mathcal{E}_{adv}^{i,B}$ on the mini-batch}
            \State {Compute loss function using Eq. \ref{loss}}
            \State {Optimize layer $i$ using the loss function}
        \EndFor
        \EndFor 
    \EndFor \\
    Trained Detector $\mathcal{D}_T$ obtained
    \State{\textcolor{violet}{/* Confidence Boundary Generation */}}
    \ForAll {i = 1 to n}
        \If{i == n}
        \State{$\mathcal{E}_{n,Th}$ $\gets$ $K^{th}$ percentile of $\Psi_{n,s}$}
        
        \Else 
        \State{Generate $i$th layer distribution, $\Psi_{i,s}$ using}
        \State{$s_{nat}$ and $\mathcal{D}_T$} 
        \State{$\mathcal{E}_{i,L}$ $\gets$ $(K-L)^{th}$ percentile of $\Psi_{i,s}$}
        \State{$\mathcal{E}_{i,U}$ $\gets$ $(K+U)^{th}$ percentile of $\Psi_{i,s}$}
        
        \EndIf
        
    \EndFor
\end{algorithmic}
\end{algorithm}
\begin{figure*}
    \centering
    \resizebox{\textwidth}{!}{
    \begin{tabular}{llllll}
         (a)& \multirow{8}{*}{\includegraphics[width=0.3\textwidth]{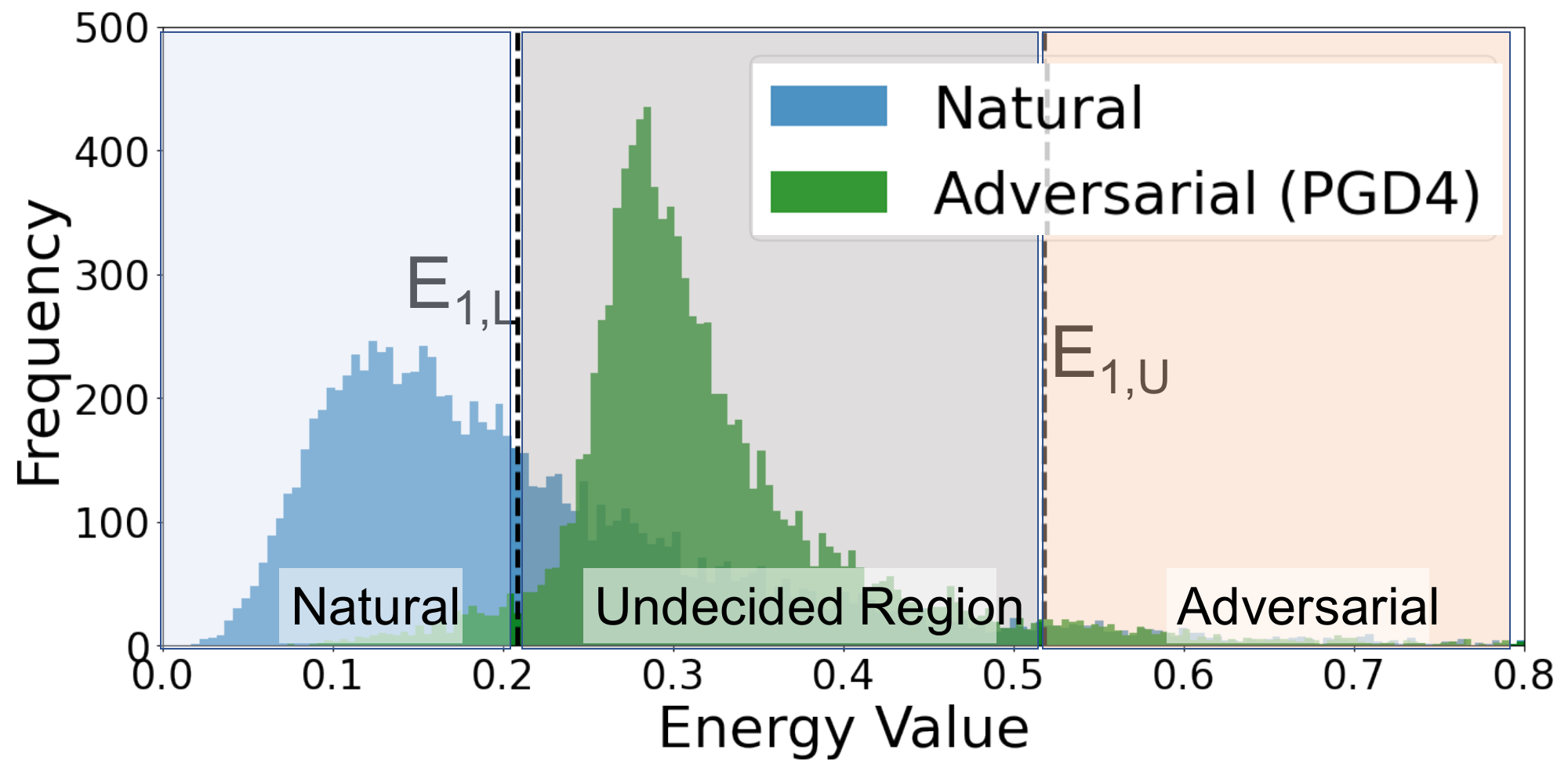}} & (b) &
         \multirow{8}{*}{\includegraphics[width=0.3\textwidth]{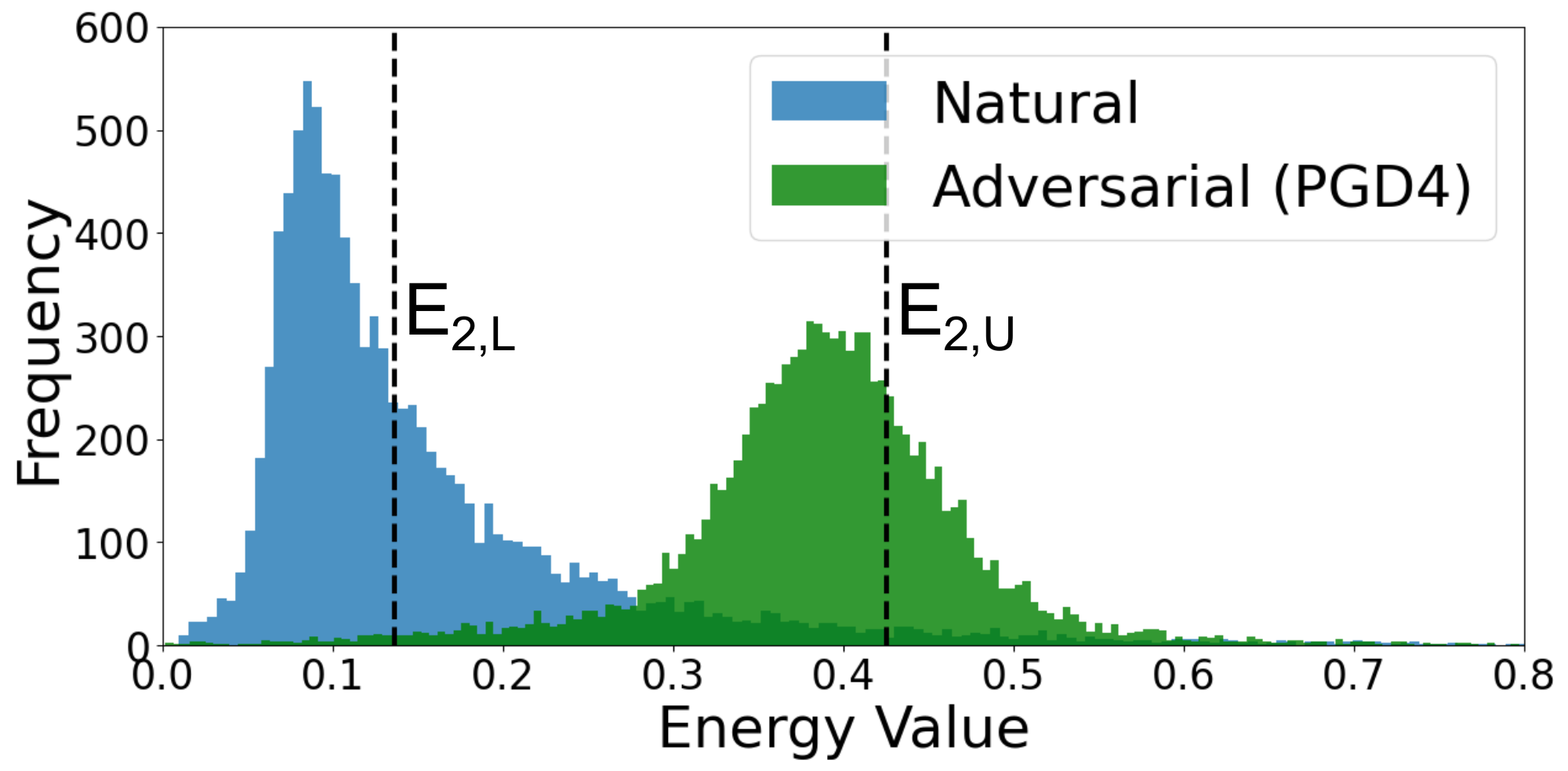}} & (c) &
         \multirow{8}{*}{\includegraphics[width=0.3\textwidth]{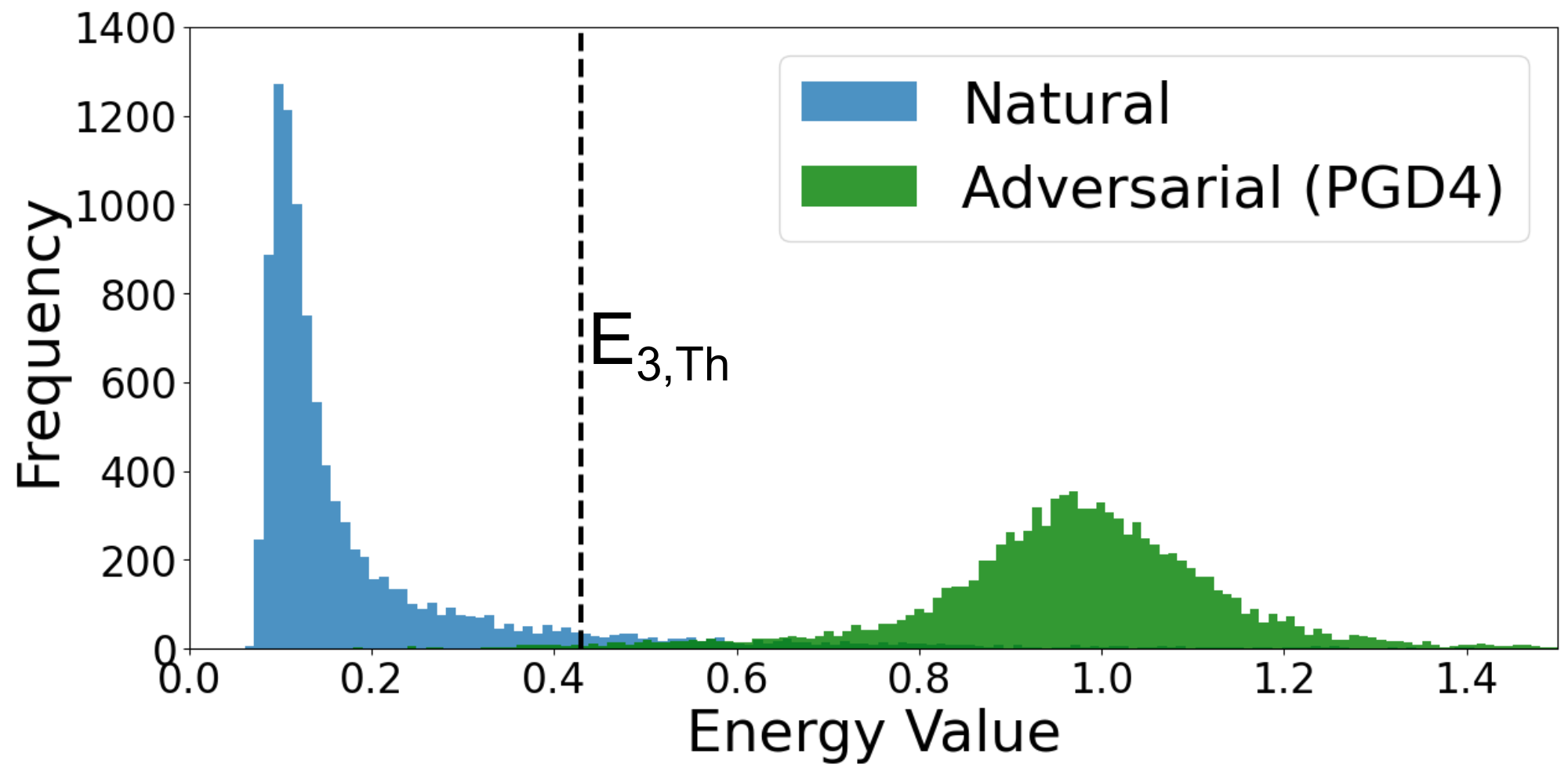}}\\
        & & & & & \\
        & & & & & \\
        & & & & & \\
        & & & & & \\
        & & & & & \\
        & & & & & \\
        & & & & & \\
          
    \end{tabular}}
    
    \caption{Plots showing the natural and adversarial \textit{energy} distributions $\mathcal{E}_{nat}^i$ and $\mathcal{E}_{adv}^i$ for layers $i$= 1, 2, 3 of a three layered detector. The confidence boundaries $\mathcal{E}_{i,L}$, $\mathcal{E}_{i,U}$ and $\mathcal{E}_{n,Th}$ shown correspond to $K$= 92, $L$= 30 and $U$=5.}
    \label{fig:energy_bound}
\end{figure*}

In QES training shown in Algorithm \ref{alg:algorithm}, we start with an $n$ layered randomly initialized detector $\mathcal{D}$. For each layer $i$ $\in$ [1,$n$], we train the detector for $N_{epoch}$ iterations. When training layer $i$, layers 1 to $i$ are quantized while layers $i+1$ to $n$ are maintained at 32-bit precision. We find that selectively quantizing the first $i$ layers leads to higher post training performance compared to quantization of all the layers. For the layer $i$, the \textit{energy signatures }($\mathcal{E}_{nat}^{i,B}$ and $\mathcal{E}_{adv}^{i,B}$) are computed over each mini-batch $B$ ($X_n^B$ and $X_a^B$) of $x_{nat}$ and $x_{adv}$, respectively. Here, $x_{nat}$ and $x_{adv}$ are the natural and adversarial datasets. The $x_{adv}$ for QES training is generated using gradient-based attacks on a separately trained DNN (\textit{i.e.,} by backpropagating through the DNN to obtain $\frac{\partial \mathcal{L}}{\partial x}$ for natural input $x$). The \textit{energy signature} for layer $i$ and mini-batch $B$ is computed using the formula, 
\begin{equation}
    \mathcal{E}^{i,B}=mean(Z_{c,h,w}^{i,B} ).
    \label{eq:energy}
\end{equation}
$Z$ is the magnitude of weighted summation outputs of layer $i$ and mini-batch $B$. Additionally, $c$, $h$ and $w$ are the number of channels, height and width of the output features, respectively. $\mathcal{E}_{nat}^{i,B}$ and $\mathcal{E}_{adv}^{i,B}$ are then used to compute the loss function given by: 
\begin{equation}
\mathcal{L}_i= y ~\mathcal{L}_{MSE} (\mathcal{E}_{nat}^{i,B},\lambda_{n}^{i})~+~(1-y)~ \mathcal{L}_{MSE} (\mathcal{E}_{adv}^{i,B},\lambda_{a}^{i}).
\label{loss}
\end{equation}
$\lambda_{n}^i$ and $\lambda_{a}^i$ are the desired natural and adversarial energies for layer $i$ and $y$ equals 1 for natural data and 0 for adversarial data. Using the loss function, we optimize the weights of the layer $i$. During optimization of layer $i$, layers 1 to $i$-1 are frozen.


\subsection{Early Detection and Exit Strategy for Energy Efficiency}
\label{early_exit}
The ``Early detection and exit" minimizes the computations and improves the energy efficiency of the adversarial detection. After QES training, we generate layer-specific sample natural \textit{energy} distributions $\Psi_{i,s}$ using the trained detector $\mathcal{D}_T$ and $s_{nat}$, as shown in Algorithm 1. Note, $s_{nat}$ is a small dataset randomly sampled from the training set. We generate layer-specific lower ($\mathcal{E}_{i,L}$) and upper ($\mathcal{E}_{i,U}$) confidence boundaries which are the $(K-L)^{th}$ and $(K+U)^{th}$ percentiles of the $\Psi_{i,s}$ distribution, respectively. The last layer has one confidence boundary, $\mathcal{E}_{n,Th}$ which is the $K^{th}$ percentile of the $\Psi_{n,s}$ distribution. Here, $K$, $U$ and $L$ are hyper-parameters and same for all the layers of the detector. 
We demonstrate the layer-specific natural and adversarial \textit{energy} distributions and confidence boundaries for PGD $\epsilon$= 4/255 (PGD4) attacks of a 3-layered trained detector on CIFAR10 data in Fig. \ref{fig:energy_bound}. For detection, the \textit{energy} $\mathcal{E}^i$ at the end of each layer $i$ is computed. As seen in Fig. \ref{fig:energy_bound}a, if the \textit{energy} is less (greater) than $\mathcal{E}_{i,L}$ ($\mathcal{E}_{i,U}$), it is classified as natural (adversarial) sample and the detection process is terminated. If the \textit{energy} value lies between $\mathcal{E}_{i,L}$ and $\mathcal{E}_{i,U}$, it is forwarded to the next layer until it is classified at the final layer. In the final layer $n$, if $\mathcal{E}^n$ is greater than $\mathcal{E}_{n,Th}$, it is classified as adversarial and vice-versa.
\section{Hardware Implementation}
\begin{figure}[h!]
    \centering
    \includegraphics[width=0.8\columnwidth]{ 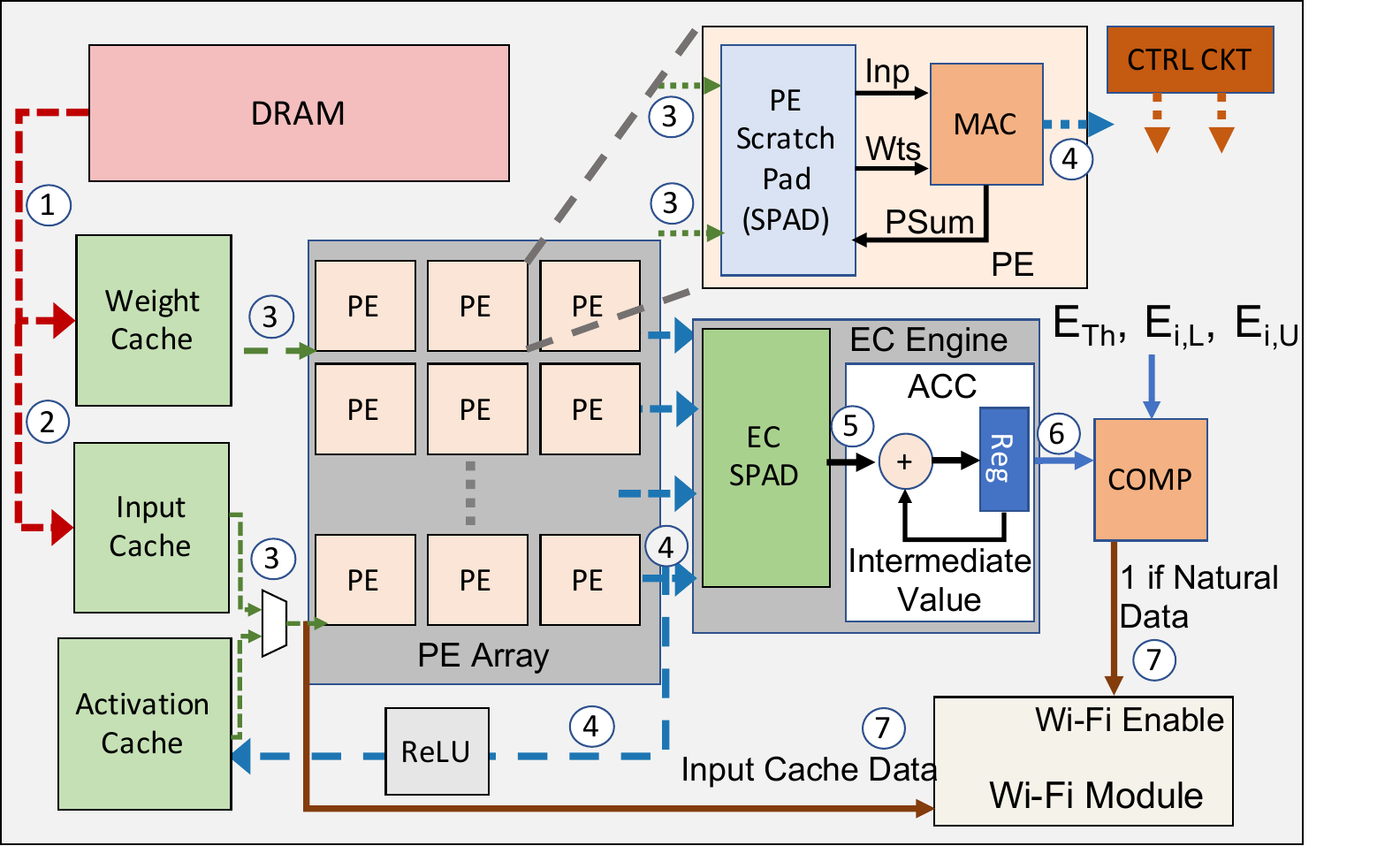}
    \caption{Hardware accelerator for an $n$ layered detector.}
    \label{fig:hardware}
\end{figure}
The QES-trained detector with ``early-detection and exit" is implemented on an Eyeriss-like \cite{chen2016eyeriss, zhang2019frequency} hardware accelerator shown in Fig. \ref{fig:hardware}. \circled{1} As the detector has a small size ($\sim$12KB), all the detector weights are fetched from the DRAM to the \textit{Weight Cache} once and reused for multiple inputs reducing DRAM accesses. \circled{2} For each input, the input data is loaded from the DRAM to the \textit{Input Cache}. \circled{3} The multiplexer selects the data from the \textit{input cache} or \textit{activation cache} depending upon the layer being processed and sends it to the \textit{PE Array}. Simultaneously, the corresponding layer weights are transferred to the \textit{PE Array}. Each PE contains weight, input and PSum scratchpads and adopt a row-stationary dataflow \cite{chen2016eyeriss} to maximize data reuse and thus improve the energy efficiency. \circled{4} The convolution outputs from each \textit{PE} is stored into the \textit{EC SPAD} and the \textit{Activation Cache} (via the \textit{ReLU} activation unit). \circled{5} A set of adder and register (\textit{Reg}) in the \textit{Energy Computation (EC) Engine} accumulate all the convolution outputs of a layer to compute the \textit{Energy} value. \circled{6} The \textit{Energy} values are compared with the layer-specific confidence boundaries ($\mathcal{E}_{i,L}$, $\mathcal{E}_{i,U}$ or $\mathcal{E}_{n,Th}$). \circled{7} If a sample is classified as natural at an early layer, the input data is fetched from the \textit{Input Cache} and directed to the WiFi module for transmission to DNN classifier on cloud and the detection ends. The adversarial detection occurs in a pipelined fashion across different images. From our analysis we find that the latency for data transmission by the WiFi module is $\sim5\times$ higher compared to the detection latency. Therefore, to ensure no data loss, we set the communication queue size in the WiFi module as 30KB ($\sim10\times$ the image size). 

\section{Experiments and Results}
\subsection{Experimental Setup}
\label{exp_setup}
\subsubsection{\textbf{``Classifier-edge" and ``Baseline" Systems}}
In RobustEdge, we implement the QES-trained detector at the edge and the classifier model is deployed on the cloud as seen in Fig. \ref{fig:motivation}c. Note, this cloud-based classifier and edge detector system will be referred to as the {``classifier-edge system"}, in the remainder of the text. Additionally, a classifier-edge system with no adversarial detection at the classifier or edge (Fig. \ref{fig:motivation}a) is considered as the {``Baseline"} for evaluation.

\subsubsection{{\textbf{Datasets}}} We use CIFAR10 (10 classes), CIFAR100 (100 classes) and TinyImagenet (200 classes) datasets. The CIFAR10 and CIFAR100 datasets have 50k training and 10k test data with each sample of size 3x32x32 pixels while TinyImagenet contains 100k training and 10k test data with each sample of size 3x64x64 pixels. 

\subsubsection{\textbf{White-box and Black-box Attack Scenarios}} 
\label{adversarial_attacks} 

When a classifier-edge system is subjected to adversarial attacks, there are two possible scenarios: white-box and black-box. In the white-box attack scenario, the adversarial attacks are generated from the same DNN model that was used to create the adversarial data during QES training. This means that the attacker has complete knowledge of the model and its parameters. On the other hand, in the black-box attack scenario, the adversarial attacks are generated from a DNN model that is different from the one used to create the adversarial data during QES training. In this scenario, the attacker has limited knowledge of the model, and this makes it harder for them to generate effective adversarial attacks. 

\subsubsection{\textbf{Adversarial Attack and Strengths}} For all the white-box and black-box attacks in our experiments, we use the following attacks with strengths as described (see Section \ref{background_adv_attacks} for details):  FGSM ($\epsilon$= 8/255) \cite{huang2017adversarial}, BIM ($\epsilon$= 8/255) \cite{kurakin2016adversarial}, DIFGSM ($\epsilon$= 8/255) \cite{xie2019improving} , MIFGSM ($\epsilon$= 8/255) \cite{dong2018boosting}, TPGD ($\epsilon$= 8/255) \cite{madry2017towards}, PGD4 (PGD with $\epsilon$= 4/255) \cite{madry2017towards} and PGD8 (PGD with $\epsilon$= 8/255) and PGD16 (PGD with $\epsilon$= 16/255), SQR ($\epsilon$= 0.3) \cite{andriushchenko2020square}, APGD ($\epsilon$= 8/255), AUTO ($\epsilon$= 16/255) \cite{croce2020reliable}, C\&W ($c$= 100, $\kappa$= 0, steps= 100) \cite{carlini2017towards} and GN ($\sigma$= 0.1).

\subsubsection{\textbf{{QES training Parameters (Algorithm \ref{alg:algorithm})}}} For all datasets, we use a 3-layered convolution neural network ($n$= 3) with the Detector $\mathcal{D}_1$ architecture shown in Fig. \ref{fig:detector_arch}. For QES training, $N_{epochs}$= 500 and requires $<2$ GPU hours on a single Nvidia RTX2080ti GPU due to the small detector size. The data batch size is 200 and [$\lambda_a^1$= 0.9, $\lambda_a^2$= 1.3, $\lambda_a^3$= 2] with $\lambda_n$ fixed at 0.1 for all the layers (Section \ref{detector_arch_sec} details the ablation studies for choice of hyperparameters). For CIFAR10 and CIFAR100, the learning rates are- 0.005, 0.002, 0.002 for optimizing layers 1, 2 and 3, respectively. For TinyImagenet, the learning rates are 0.003, 0.002, 0.002. For each dataset scenario, the QES detector is trained with $x_{adv}$ created using a separately trained VGG19 for CIFAR10/CIFAR100, and ResNet18 for TinyImagenet. In all experiments, $x_{adv}$ are PGD $\epsilon$ = 8/255 attack based inputs. The $s_{nat}$ dataset has 1000 images randomly sampled from the training set. {The values of $L$, $U$ and $K$ in all experiments are 30, 5 and 92, respectively} (see Fig. \ref{fig:klu_params}). We use Pytorch v1.5.1 for all experiments.   
\subsubsection{{\textbf{Hardware Evaluation Setup}}}
\begin{table}[h!]
    \centering
    \resizebox{\columnwidth}{!}{
    \begin{tabular}{|c|ccccc|}\hline
        Technology  &  \multicolumn{5}{c|}{45nm CMOS} \\ \hline
        DRAM & \multicolumn{5}{c|}{64MB} \\ \hline
        Input, Weight, Activation Cache &  \multicolumn{5}{c|}{32KB SRAM} \\ \hline
        PE \& EC SPAD & \multicolumn{5}{c|}{1KB Register} \\ \hline
       PE Array Size & \multicolumn{5}{c|}{32} \\ \hline
       \multicolumn{6}{|c|}{\textbf{Energy Values}} \\ \hline
        $Precision$ & 4-bits & 6-bits & 8-bits & 12-bits & 16-bits  \\ \hline
        $E_{MAC}$ / Op (pJ) & 0.0575 &  0.129 & 0.23  &   0.52 &  0.92 \\ \hline
        $E_{ACC}$ / Op (pJ) & 0.017 &  0.038 & 0.07   &  0.15 &  0.27 \\
        \hline
        $E_{DRAM}$ / Read (16-bits) (pJ) & 184 & 184 & 184 & 184 & 184 \\ \hline
        $E_{CACHE}$ / Read (16-bits) (pJ) & 10 & 10 & 10 & 10 & 10 \\ \hline
        $E_{SPAD}$ / Read (16-bits) (pJ) & 1.7 & 1.7 & 1.7 & 1.7 & 1.7 \\ \hline
        $E_{transmit}$ / Image (mJ) & \multicolumn{5}{c|}{CIFAR10, CIFAR100- 6.8,       TinyImagenet- 13.6}\\ \hline
    \end{tabular}}
    \caption{Table showing parameters for the hardware accelerator used for implementing the detector.}
    \label{tab:systolic_params} 
\end{table}
The classifier-edge system's energy includes natural and adversarial data transmission energy from edge to classifier ($E_{T,N}$ and $E_{T,A}$) and the adversarial detection energy $E_D$ as shown in Eq. \ref{total_energy}.
\begin{equation}
E = E_{T,N} + E_{T,A} + E_D
\label{total_energy}
\end{equation}
\begin{equation}
E_{T,N} = p\times N_{nat} \times \frac{E_{transmit}}{Image}, ~ E_{T,A} = q\times N_{adv}\times \frac{E_{transmit}}{Image}
\label{eta_etn}
\end{equation}
\begin{multline}
    E_{D} = (N_{nat}+N_{adv})\times (\frac{E_{DRAM}}{Read}\times R_D + \frac{E_{CACHE}}{Read}\times R_C\\ + \frac{E_{SPAD}}{Read}\times R_S + \frac{E_{MAC}}{Op}\times N_{MAC} + \frac{E_{ACC}}{Op}\times N_{ACC})
    \label{Ed_rob}
\end{multline}
    
In Eq. \ref{eta_etn}, $N_{nat}$ and $N_{adv}$ are the number of natural and adversarial samples. $E_{transmit}$ is the transmission energy per image to send data from edge to the classifier. $p$ and $q$ are the fraction of natural and adversarial samples, respectively transmitted to the classifier. Eq. \ref{Ed_rob} shows that the $E_D$ includes the total DRAM, Cache, SPAD, MAC and ACC component  energies computed over all the natural and adversarial inputs. An ideal detector should have $p$=1 and $q$=0 (or $E_{T,A}$= 0) at low detection cost $E_D$. 

The QES-trained quantized detector (see Fig. \ref{fig:motivation}c) is implemented using the hardware platform (Fig. \ref{fig:hardware}) with parameters shown in Table \ref{tab:systolic_params}. All energy evaluations in Table \ref{tab:systolic_params} are based on SPICE simulations using the Cadence Virtuoso platform. For all experiments, the transmission energy per image, $E_{transmit}$ is computed using the Wi-Fi power model proposed in \cite{huang2012close} at a transfer rate of 100Mbps. Note, the comparator and ReLU units' energies are not shown as they are negligible compared to the other components.

\subsubsection{\textbf{Performance Evaluation Metrics}}
For evaluating detector performance, we use the metric \textbf{AUC and F1 score} \cite{yin2019gat}: High {AUC and F1 scores (close to 1) signifies} a high true positive and low false positive rate and hence a good detector. For evaluating the Classifier-Edge system's performance, we use 1) \textbf{Error} \cite{yin2019gat}: which is the percentage of total adversarial inputs that are undetected and misclassified by the classifier model. 2) \textbf{Accuracy} \cite{yin2019gat}: which is the percentage of natural samples that are identified correctly by the detector and the classifier. High \textit{accuracy} and low \textit{error} imply higher robustness. 
\subsection{Adversarial Robustness and Energy-efficiency Results}
\label{rob_energy_eff}
\begin{table*}
\centering
\resizebox{\textwidth}{!}{
\begin{tabular}{c}

    \begin{tabular}{|l|cc|c|c|c|c|c|c|c|c|c|c|}\hline
         \textbf{CIFAR100} &\multicolumn{2}{c}{Attacks $\rightarrow$}  &\: FGSM &\: PGD4 & \: PGD8 &\: PGD16 &\: TPGD &\: DIFGSM &\: MIFGSM & \: C\&W & \: AUTO & \: GN  \\
         \hline
         \multirow{3}{*}{\begin{tabular}{l} \textbf{16b Edge Detector}\\ Classifier: VGG19 \\ \:\end{tabular}} & \multicolumn{2}{c|}{Error R / Error B} &\: 0 / 82 &\: 0 / 95 & \: 0 / 96 &\: 0 / 95 &\:0 / 97 &\: 0 / 80 &\: 0 / 88 &\: 6 / 89 &\: 13 / 69 &\: 0 / 77 \\ \cline{4-13}
         & \multicolumn{2}{c|}{Accuracy R / Accuracy B} & 59 / 62 & 59 / 62 & 59 / 62 & 59 / 62 & 59 / 62 & 59 / 62 & 59 / 62 & 59 / 62 & 59 / 62 & 59 / 62\\ \cline{2-13}
         & \multicolumn{2}{c|}{16b Detector's AUC Score} & 0.972 & \textbf{0.965} & \textbf{0.978} & \textbf{0.98} & \textbf{0.976} & \textbf{0.971} & 0.975 & \textbf{0.8} & \textbf{0.8} & 0.973 \\ \cline{2-13}

         & \multicolumn{2}{c|}{16b Detector's F1 Score} & 0.962 & 0.953 & {0.962} & 0.957 &  0.956 & {0.957} & {0.963} & {0.721} &{0.722} &\textbf{0.961} \\ \hline
         
         \multirow{3}{*}{\begin{tabular}{l} \textbf{12b Edge Detector}\\ Classifier: VGG19 \\ \:\end{tabular}} & \multicolumn{2}{c|}{Error R / Error B} &\: 0 / 82 &\: 0 / 95 & \: 0 / 96 & \: 0 / 95 &\:0 / 97 &\: 0 / 80 &\: 0 / 88 &\: 8 / 85 &\: 15 / 69 &\: 0 / 77 \\ \cline{4-13}
         & \multicolumn{2}{c|}{Accuracy R / Accuracy B} & 59 / 62 & 59 / 62 & 59 / 62 & 59 / 62 & 59 / 62 & 59 / 62 & 59 / 62 &59 / 62 &59 / 62 & 59 / 62\\ \cline{2-13}
         & \multicolumn{2}{c|}{12b Detector's AUC Score} & \textbf{0.98} & 0.963 & {0.978} & 0.972 &  0.975 & {0.971} & \textbf{0.98} & {0.79} &{0.79} &\textbf{0.98} \\ \cline{2-13}

         & \multicolumn{2}{c|}{12b Detector's F1 Score} & 0.962 & 0.953 & {0.962} & 0.957 &  0.956 & {0.957} & {0.963} & {0.721} &{0.722} &\textbf{0.961} \\ \hline
    \end{tabular}\\ \\
    
    \begin{tabular}{|l|cc|c|c|c|c|c|c|c|c|c|c|}\hline
         \textbf{TinyImagenet} &\multicolumn{2}{c}{Attacks $\rightarrow$}  &\: FGSM &\: PGD4 & \: PGD8 & \: PGD16 &\: TPGD &\: DIFGSM &\: MIFGSM &\: C\&W &\: AUTO &\: GN  \\
         \hline
         \multirow{3}{*}{\begin{tabular}{l} \textbf{16b Edge Detector} \\ Classifier: ResNet18 \\ \:\end{tabular}} & \multicolumn{2}{c|}{Error R / Error B} &\: 0 / 89 &\: 0 / 95 & \: 0 / 95 & \: 0 / 97 &\: 0 / 91 &\: 0 / 82 &\: 0 / 80 &\: 10 / 94 &\: 18 / 77 &\: 0 / 71 \\ \cline{4-13}
         & \multicolumn{2}{c|}{Accuracy R / Accuracy B} & 50 / 53 & 50 / 53 & 50 / 53 & 50 / 53 & 50 / 53 & 50 / 53 & 50 / 53 &50 / 53 & 50 / 53 & 50 / 53\\ \cline{2-13}
         & \multicolumn{2}{c|}{16b Detector's AUC Score} & 0.978 & \textbf{0.973} & \textbf{0.977} & 0.978 & \textbf{0.975} & \textbf{0.977} & 0.978 &\textbf{0.7} & \textbf{0.71} & 0.975 \\ \cline{2-13}

         & \multicolumn{2}{c|}{16b Detector's F1 Score} & 0.955 & 0.936 & {0.954} & 0.956 &  0.958 & {0.954} & {0.954} & {0.628} &{0.611} & {0.956} \\ \hline
         
         \multirow{3}{*}{\begin{tabular}{l} \textbf{12b Edge Detector} \\ Classifier: ResNet18 \\ \:\end{tabular}} & \multicolumn{2}{c|}{Error R / Error B} &\: 0 / 89 &\: 0 / 95 & \: 0 / 95 &\: 0 / 97 &\: 0 / 91 &\: 0 / 82 &\: 0 / 80 &\: 15 / 94 & \: 20 / 77 &\: 0 / 71 \\ \cline{4-13}
         & \multicolumn{2}{c|}{Accuracy R / Accuracy B} & 50 / 53 & 50 / 53 & 50 / 53 & 50 / 53 & 50 / 53 & 50 / 53 & 50 / 53 &50 / 53 &50 / 53 &50 / 53\\ \cline{2-13}
         & \multicolumn{2}{c|}{12b Detector's AUC Score} & \textbf{0.98} & {0.973} & 0.973 & {0.98} & 0.972 & 0.973 & \textbf{0.98} & {0.67} & {0.67} & \textbf{0.98} \\ \cline{2-13}

         & \multicolumn{2}{c|}{12b Detector's F1 Score} & 0.955 & 0.936 & {0.954} & 0.956 &  0.958 & {0.954} & {0.954} & {0.628} &{0.611} & {0.956} \\ \hline
    \end{tabular}

\end{tabular}}
\caption{Table comparing the \textit{Error}, \textit{Accuracy} {AUC and F1 scores} of RobustEdge-based classifer-edge systems with 16b and 12b detectors across different adversarial attacks for CIFAR100 and TinyImagenet datasets against the ``Baseline". ``R" and ``B" denote RobustEdge and ``Baseline" systems, respectively. We use classifier VGG19 model for CIFAR100 and ResNet18 for TinyImagenet trained on natural data using standard SGD. Note, the accuracy remains constant for all adversarial attacks as it is computed for natural samples. For a specific attack, the highest AUC score is highlighted in bold. }
\label{tab:diff_attacks_err}
\end{table*}
As seen in Table \ref{tab:diff_attacks_err}, the RobustEdge based classifier-edge systems with 16b and 12b QES trained detectors achieve \textit{Error}= 0 and near iso-\textit{Accuracy} compared to the ``Baseline" (see Section \ref{exp_setup}). Here, the evaluations are based on white-box attacks (refer Section \ref{adversarial_attacks}). For both ``Baseline" and RobustEdge scenarios, the classifier on the cloud is a VGG19 model for CIFAR100 and ResNet18 model for TinyImagenet trained only on natural data using SGD algorithm. {Note, the ``Baseline" has $p= 1$, $q= 1$ and $E_D= 0$} and therefore incurs high \textit{Error}. RobustEdge achieves significantly {high AUC and F1 scores} ($>$0.9 for gradient and $>0.7$ for score-based attacks) which significantly reduces the \textit{Error} rate while maintaining near iso-\textit{Accuracy} performance. Adversarial detection at the edge minimizes the communication of adversarial samples ($q$= 0) while maximizing the communication of natural samples to the classifier (p$\sim$1).
\begin{figure}[h!]
    \centering
    \resizebox{1\columnwidth}{!}{
    \begin{tabular}{cc}
          {\includegraphics[width=0.5\columnwidth]{ 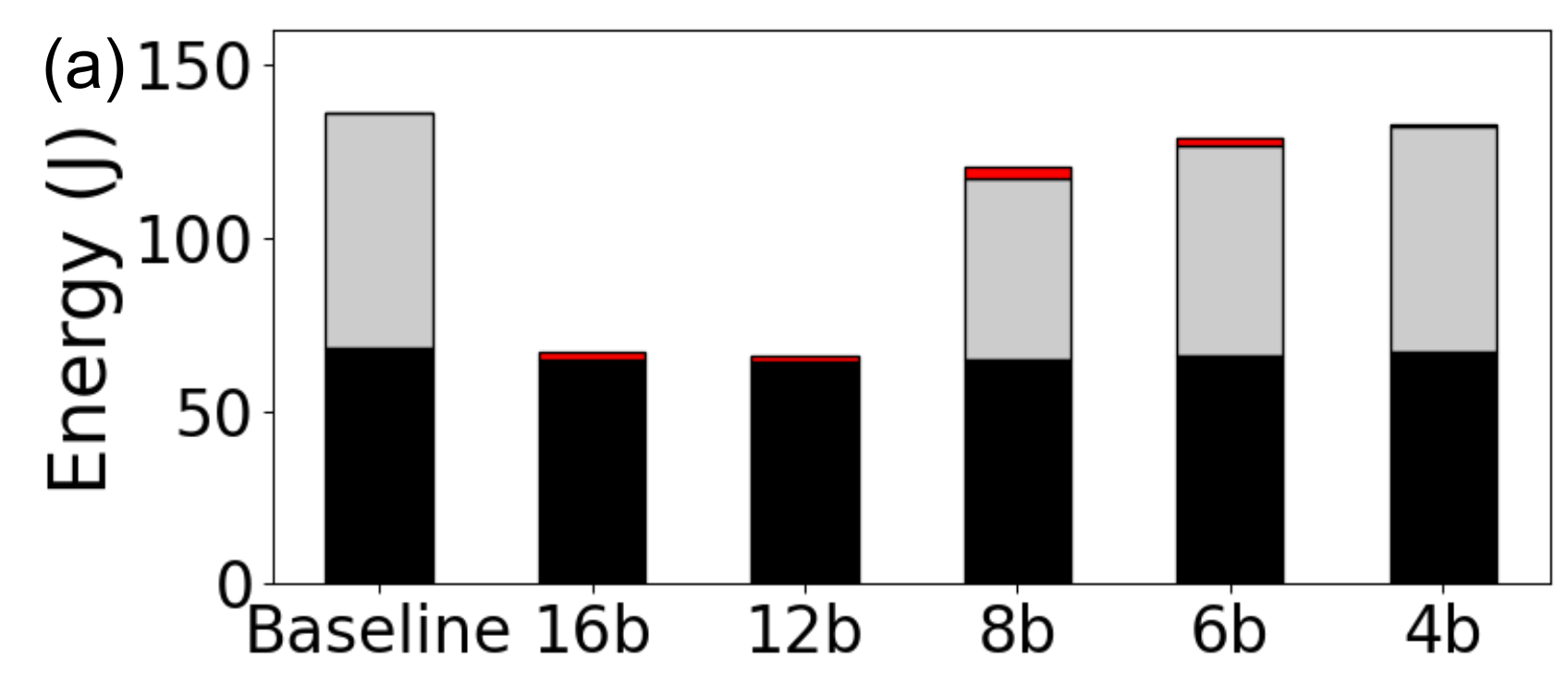}\hspace{-2mm}}  &
          {\includegraphics[width=0.5\columnwidth]{ 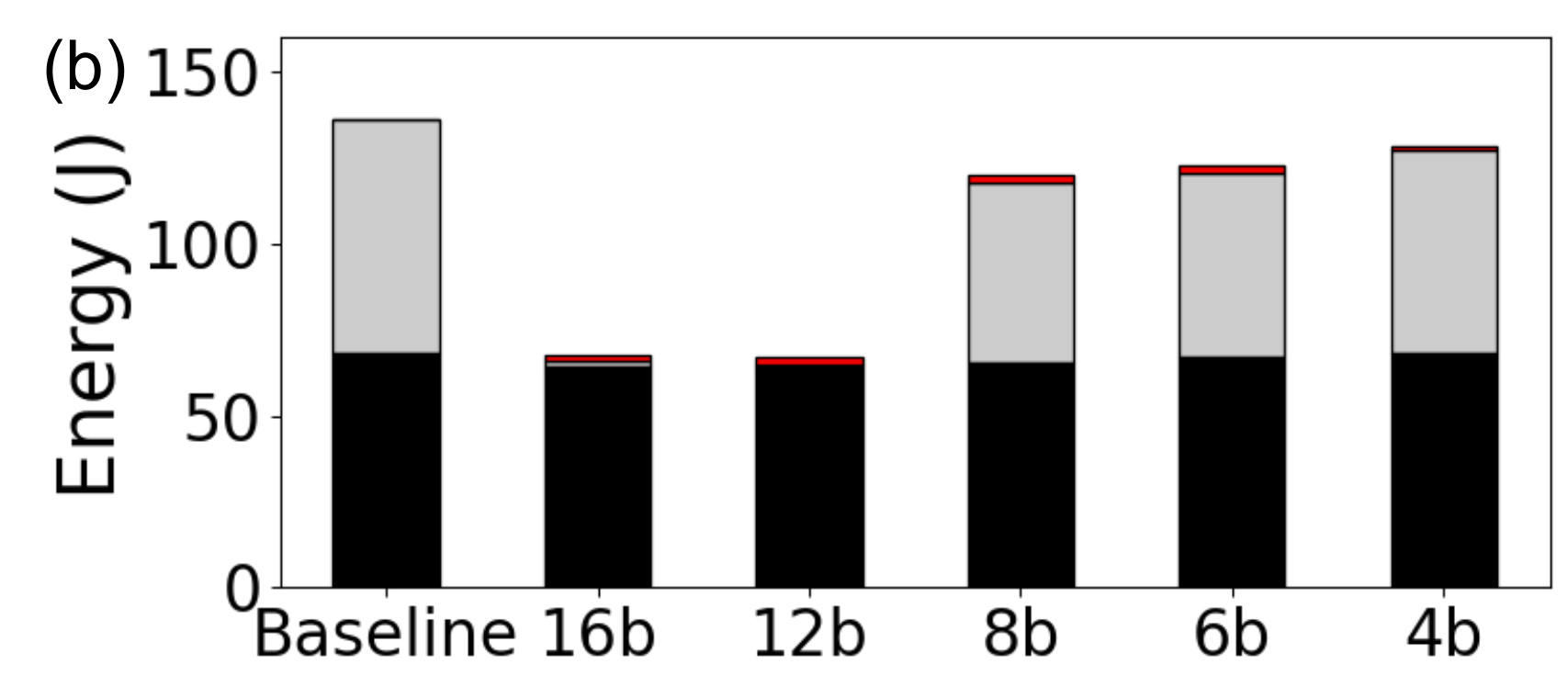}}  \\

         {\includegraphics[width=0.5\columnwidth]{ 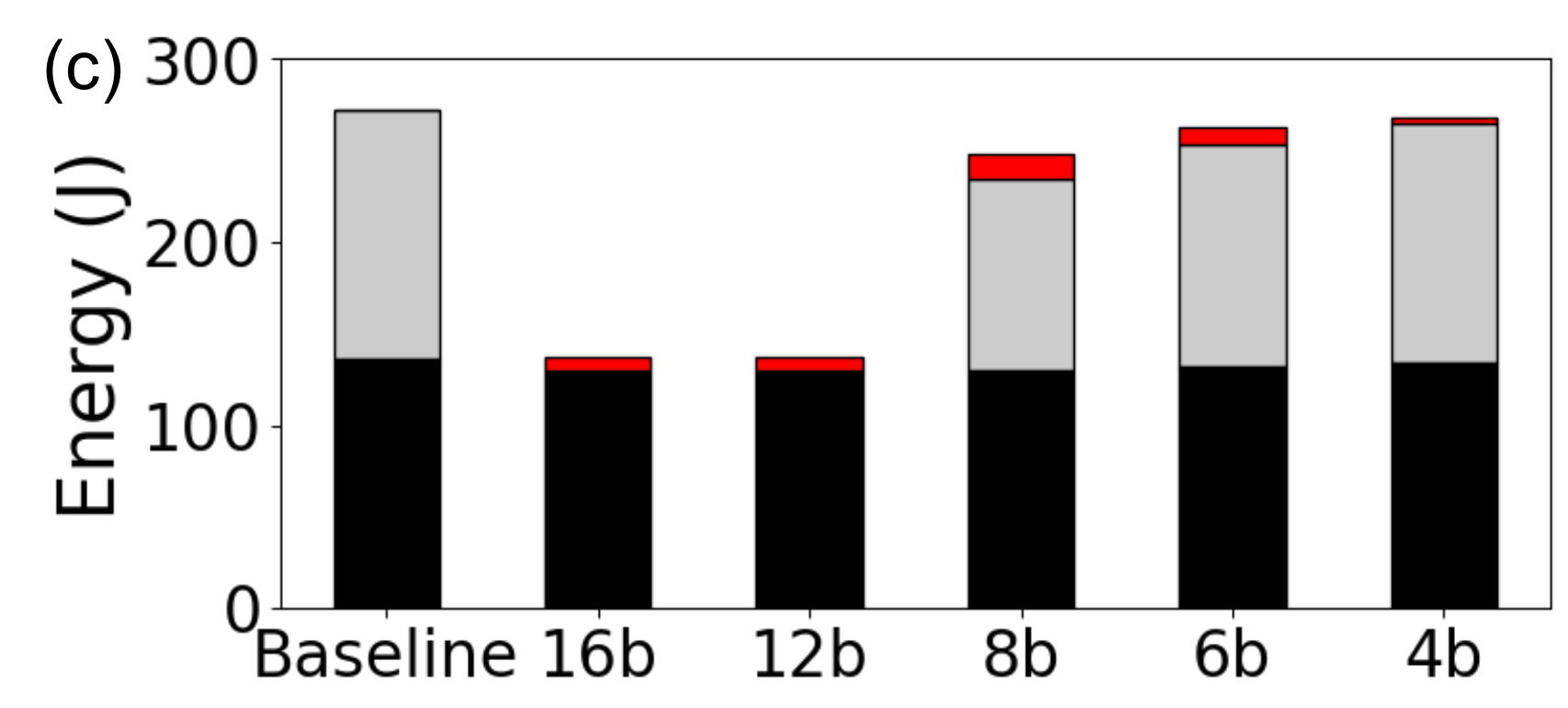}}  &
         {\includegraphics[width=0.5\columnwidth]{ 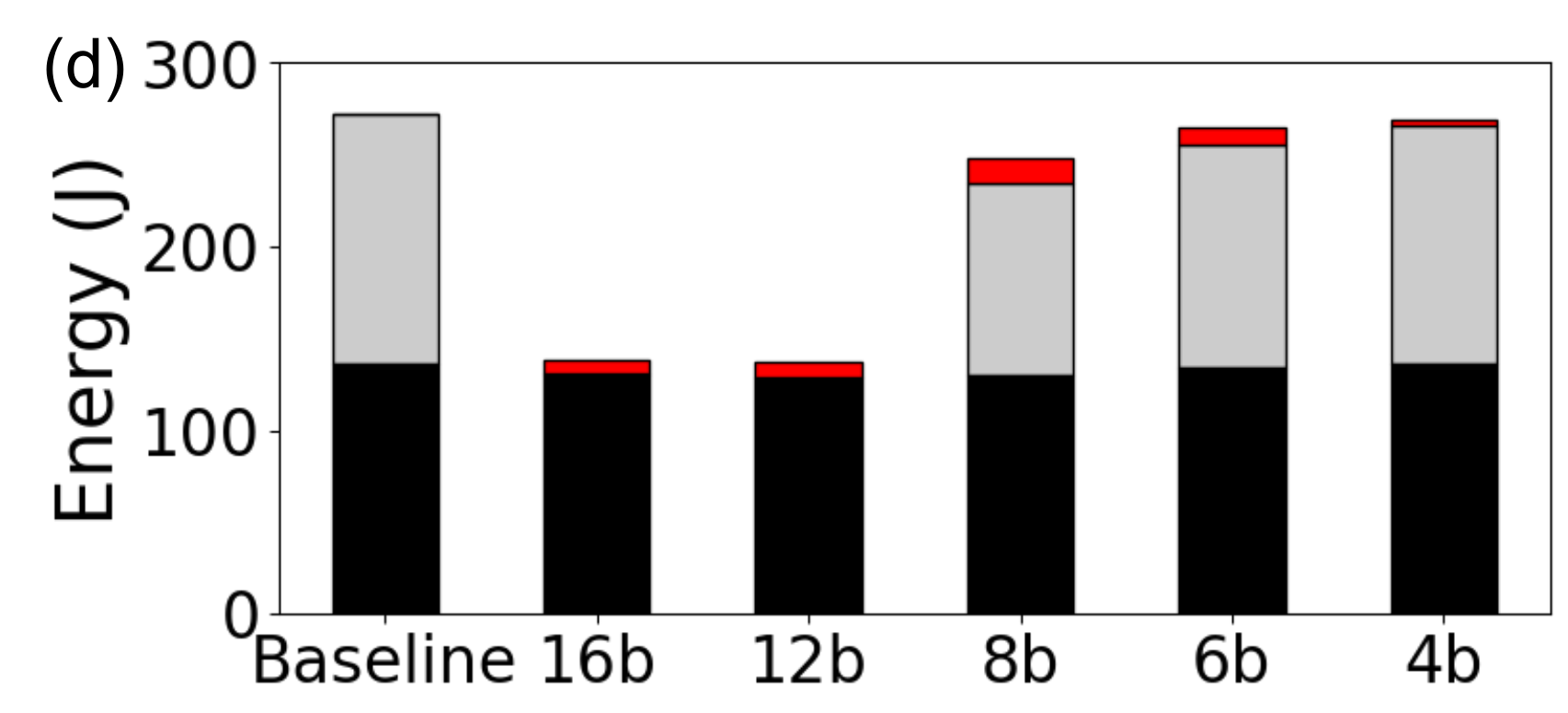}}  \\

         \multicolumn{2}{c}{\includegraphics[width=0.4\columnwidth]{ 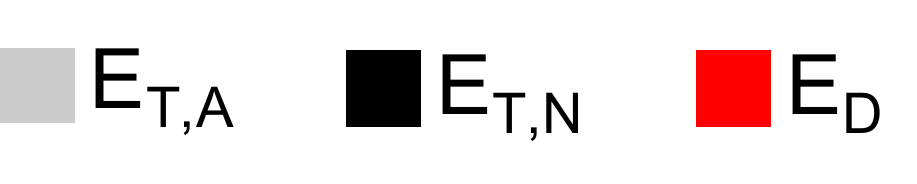}}
         
    \end{tabular}}
    \caption{Figure showing the $E_{T,A}$, $E_{T,N}$ and $E_D$ for the ``Baseline" and RobustEdge-based classifier-edge systems with different quantized detectors implemented at the edge. Here, the detectors are implemented on the 45nm CMOS hardware accelerator shown in Fig. \ref{fig:hardware}. Energy results correspond to (a) PGD16 attacks CIFAR100 (b) PGD4 attacks CIFAR100 (c) PGD16 attacks TinyImagenet (d) PGD4 attacks TinyImagenet datasets. In all cases, $N_{nat} = N_{adv} = 10k$. $E_D$ is computed using Eq. \ref{Ed_rob} with parameters in Table \ref{tab:systolic_params}}
    \label{fig:eta_etn_ed_c100_tiny}
\end{figure}

From Fig. \ref{fig:eta_etn_ed_c100_tiny}, we find that due to high adversarial detection, the RobustEdge-based classifier-edge systems with 16b and 12b edge detectors achieve $E_{T,A}$= 0 (as $q=0$) and $E_{T,N}$ close to the $E_{T,N}$ of the ``Baseline" (as $p\sim$1). Additionally, due to ``early-detection and exit", the detection energy $E_D$ is significantly low. For lower bit precision detectors (8b, 6b and 4b) the system incurs higher $E_{T,A}$ and $E_D$ compared to 16b and 12b detectors due to poor adversarial detection and less frequent ``early-detection and exit" (leading to higher computations). 
\subsection{Performance under Black-Box Attacks}
\label{bb_attack}

\begin{table*}
\centering
\resizebox{0.7\textwidth}{!}{
\begin{tabular}{lcc|cc|cc}
& \multicolumn{2}{c}{Scenario 1} & \multicolumn{2}{c}{Scenario 2} & \multicolumn{2}{c}{Scenario 3}\\ \hline
$DNN_{QES} \rightarrow$ \:& \multicolumn{2}{c|}{MobileNet-V2}& \multicolumn{2}{c|}{ ResNet18} & \multicolumn{2}{c}{VGG16} \\ \hline
 $DNN_{Test} \rightarrow$ \:& \: ResNet18 &\: VGG16 \:& \: VGG16 & \:MobileNet-V2 \:&\: ResNet18 &\: MobileNet-V2  \\
\hline
\multirow{14}{*}{\begin{tabular}{l|}
     FGSM \cite{kurakin2016adversarial} \:\\ BIM \cite{kurakin2016adversarial} \:\\ PGD8 \cite{madry2017towards} \: \\ PGD4 \cite{madry2017towards} \: \\PGDL2 \cite{madry2017towards} \: \\ FFGSM \cite{wong2020fast}  \:\\ TPGD \cite{zhang2019theoretically} \:\\ MIFGSM \cite{dong2018boosting} \:\\ DIFGSM \cite{xie2019improving}\:\\ C\&W  \cite{carlini2017towards} \: \\ SQR \cite{andriushchenko2020square} \\ APGD \cite{croce2020reliable} \:\\ AUTO \cite{croce2020reliable} \:\\ GN \:\end{tabular}} & \multirow{14}{*}{\begin{tabular}{c}0.97\\ 0.97\\ 0.97\\ 0.97\\ 0.95\\ 0.97\\ 0.97\\ 0.97\\ 0.97\\ 0.71\\ 0.71\\ 0.68\\ 0.71\\ 0.97 \end{tabular}} & \multirow{14}{*}{\begin{tabular}{c}0.97\\ 0.97\\ 0.97\\ 0.97\\ 0.97\\ 0.97\\ 0.97\\ 0.97\\ 0.97\\ 0.67\\ 0.67\\ 0.67\\ 0.67\\ 0.97 \end{tabular}} & \:\: \multirow{14}{*}{\begin{tabular}{c}0.97\\ 0.97\\ 0.97\\ 0.97\\ 0.97\\ 0.97\\ 0.97\\ 0.97\\ 0.97\\ 0.67\\ 0.67\\ 0.67\\ 0.67\\ 0.97 \end{tabular}} & \multirow{14}{*}{\begin{tabular}{c}0.97\\ 0.97\\ 0.97\\ 0.97\\ 0.95\\ 0.97\\ 0.97\\ 0.97\\ 0.97\\ 0.71\\ 0.71\\ 0.68\\ 0.71\\ 0.97 \end{tabular}} & \multirow{14}{*}{\begin{tabular}{c}0.98 \\ 0.98 \\ 0.98 \\ 0.97 \\ 0.92 \\ 0.98 \\ 0.98 \\ 0.98 \\ 0.98 \\ 0.75 \\ 0.75 \\ 0.74 \\ 0.75 \\ 0.98 \end{tabular}} & \multirow{14}{*}{\begin{tabular}{c}0.97\\  0.97\\  0.97\\ 0.97\\ 0.95\\ 0.97\\ 0.97\\ 0.97\\ 0.97\\ 0.73\\ 0.72\\ 0.68\\ 0.7\\ 0.98 \end{tabular}} \\ 
 &   &  & \:\:  &  \\ 
 &   &  & \:\:  & \\ 
 &   &  & \:\:  & \\ 
 &   & & \:\:  & \\ 
 &   &  & \:\: & \\ 
 &   &  & \:\:  & \\ 
 &   &  & \:\:  &\\ 
 &   &  & \:\:  & \\ 
 &   &  & \:\:  & \\ 
 &   &  & \:\:  & \\ 
 &  &  & \:\:&  \\ 
 &   &  & \:\:  &\\ 
 &   &  & \:\: & \\ 
  \hline 
\end{tabular}}
\caption{Table showing the AUC scores for different detectors trained on the TinyImagenet dataset. $DNN_{QES}$ and $DNN_{Test}$ are the DNN models used for generating adversarial attacks during QES training and test times.}
\label{tab:model_agnostic_tiny}
\end{table*}

QES-trained detectors achieve high detection performance under black-box scenarios. In Table \ref{tab:model_agnostic_tiny} we test 16-bit QES-trained detectors under three black-box attack scenarios for the TinyImagenet dataset. In Scenario 1, the detector is created using adversarial samples generated from $DNN_{QES}$= MobileNet-V2 model (the MobileNet-V2 model is trained on natural data using SGD) and tested on adversarial data generated using $DNN_{Test}$= ResNet18 and VGG16 models (trained on natural data using SGD). Similarly, for scenarios 2 and 3, the $DNN_{QES}$ are ResNet18 and VGG16, respectively with $DNN_{Test}$ models chosen accordingly. Interestingly, we observe that the QES-trained detectors have high AUC scores across different gradient and score-based adversarial attacks irrespective of the $DNN_{QES}$ or $DNN_{Test}$ models. Similar observations are made on the CIFAR10 and CIFAR100 datasets. 

\subsection{Comparison with Prior Works}
\label{comparison}

\begin{table}[t]
\centering
\resizebox{1\columnwidth}{!}{%
\begin{tabular}{|l|c|c|c|c|c|}
\hline
{Work} & {Dataset} & PGD4 & PGD16 & {\#Params} & {\#Ops}\\ \hline
Metzen et al. \cite{metzen2017detecting} & CIFAR10 & 0.96 & N-R & 312x & 17x \\ \hline
Moitra et al. \cite{moitra2021detectx} & CIFAR10 & 0.88 & 0.895 & 0.3x & 0.29x \\ \hline
Sterneck et al. \cite{sterneck2021noise} & CIFAR10 & 0.998 & 1 & 98x & 6.4x\\ \hline
Xu et al. \cite{xu2017feature} & CIFAR10 & 0.505 &  N-R & 338x & 82x\\ \hline
\textbf{RobustEdge} & \textbf{CIFAR10} & \textbf{0.98}  & \textbf{0.98} & \textbf{1x} & \textbf{1x}\\ 
\hline
Sterneck et al. \cite{sterneck2021noise} & CIFAR100 & 0.99 & 1 & 98x & 6.4x\\ \hline
Moitra et al. \cite{moitra2021detectx} & CIFAR100 & 0.64 & 0.98 & 0.3x & 0.29x\\ \hline
\textbf{RobustEdge} & \textbf{CIFAR100} & \textbf{0.97} & \textbf{0.98}& \textbf{1x} & \textbf{1x}\\ 
\hline
Moitra et al. \cite{moitra2021detectx} & TinyImagenet & 0.56 & 0.65 & 0.3x & 0.29x\\ \hline
\textbf{RobustEdge} & \textbf{TinyImagenet} & \textbf{0.98} & \textbf{0.98} & \textbf{1x} & \textbf{1x}\\ 
\hline
\end{tabular}%
}
\caption{{Table comparing the AUC scores, parameter and operation footprint of 16b RobustEdge detector with prior state-of-the-art detection works. The number of parameters and operations are normalized with respect to RobustEdge. ``N-R" denotes ``Not-Reported" AUC scores in prior works.}}
\label{tab:comparison_pw}
\end{table}

Prior detection works such as Xu et al \cite{xu2017feature} use the outputs of multiple trained classifier DNNs to perform adversarial detection. Recently, works by Metzen et al. \cite{metzen2017detecting} and Sterneck et al. \cite{sterneck2021noise} propose to train binary detector networks on intermediate layer activations of the DNN classifier for adversarial detection. Another work, DetectX \cite{moitra2021detectx} proposes to train the first convolution layer of a classifier for adversarial detection using current signatures in analog crossbar arrays. All prior detection works have the following characteristics: \textbf{1) The detector models are attached to the classifier model and hence detection always occurs at the classifier (high $E_{T,A}$). 2) Both the classifier model and detector execute simultaneously to perform adversarial detection (high $E_{D}$).} This makes them unsuitable for low power edge implementations.

In RobustEdge, the detector is significantly small and is detached from the classifier model. Under white-box attack scenarios, as seen in Table \ref{tab:comparison_pw}, {RobustEdge achieves similar or higher detection performance at extremely low parameter and operation footprint} ($100\times$ and $10\times$ lesser parameters and operations, respectively compared to \cite{xu2017feature}, \cite{metzen2017detecting} and \cite{sterneck2021noise}) compared to prior detection works. Note, although RobustEdge's parameter and operation footprint is slightly higher compared to Moitra et al. \cite{moitra2021detectx}, we achieve significantly higher detection performance (Table \ref{tab:comparison_pw}).

\begin{figure}
    \centering
    \resizebox{0.8\columnwidth}{!}{
    \begin{tabular}{c}
         \includegraphics[width=0.8\columnwidth]{ 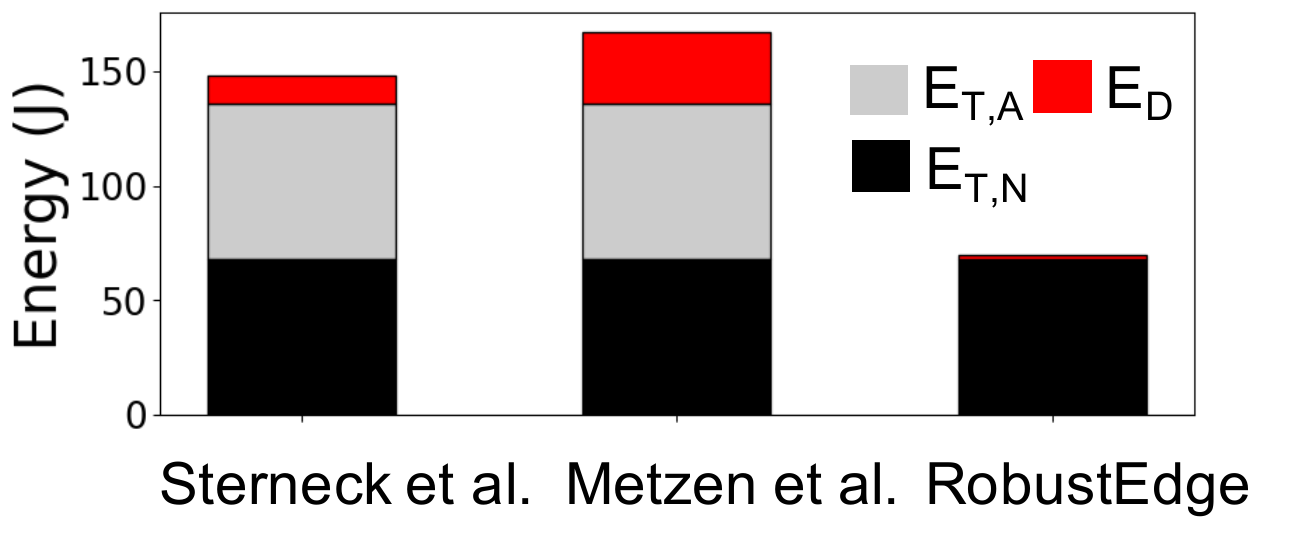}\vspace{-2mm} \\ (a) \\
         \includegraphics[width=0.8\columnwidth]{ 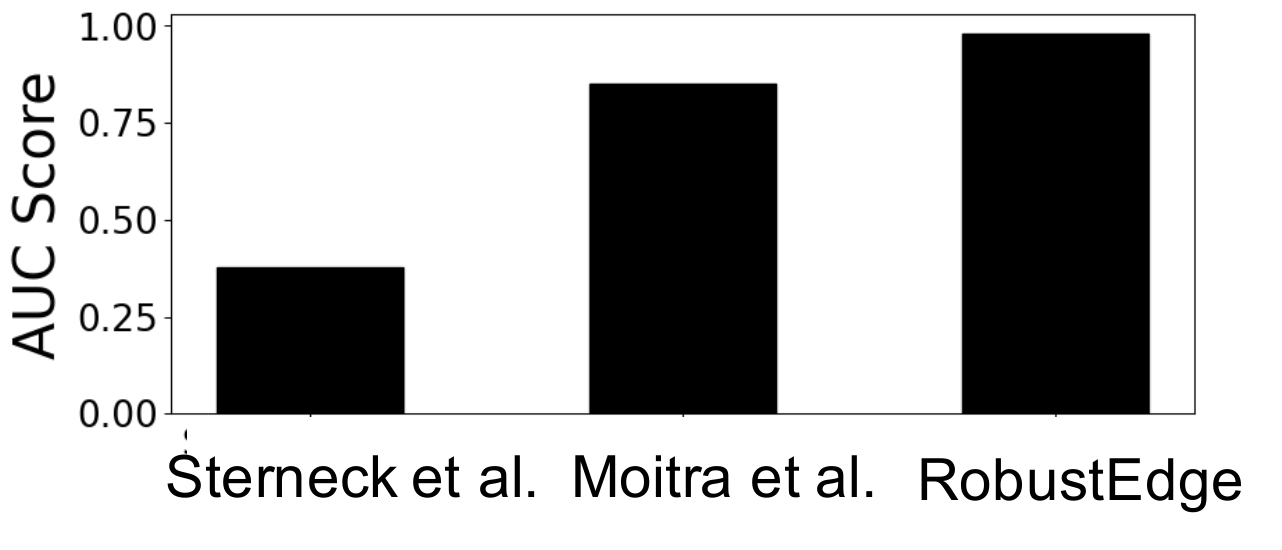}\vspace{-2mm} \\ (b)
    \end{tabular}}

    \caption{(a) $E_{T,A}$, $E_{T,N}$ and $E_D$ comparison of RobustEdge with prior works \cite{metzen2017detecting, sterneck2021noise}. $E_{T,A}$, $E_{T,N}$ and $E_D$ are computed under PGD8 attacks, $N_{nat}$=$N_{adv}$=10k for CIFAR100 dataset. (b) Detection AUC scores for different prior works \cite{sterneck2021noise, moitra2021detectx} under black-box PGD8 attacks for the CIFAR100 dataset.}
    \label{fig:inf_training_energy_pw} 
\end{figure}
\begin{figure*}[t]
    \centering
    \begin{tabular}{c}
         \includegraphics[width=0.8\textwidth]{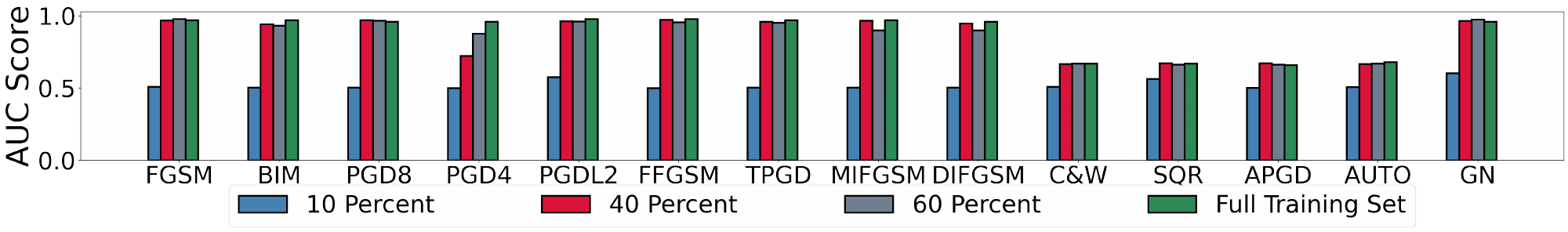} 
         
    \end{tabular}

    \caption{AUC scores corresponding to different adversarial attacks for detectors created using QES training with 10\% , 40\% and 60\% of the TinyImagenet dataset.}
    \label{fig:lim_data}
    
\end{figure*}
Fig. \ref{fig:inf_training_energy_pw}a compares the $E_{T,A}$, $E_{T,N}$ and $E_D$ of a RobustEdge-based classifier-edge system having a 16b edge detector with prior works- Sterneck et al \cite{sterneck2021noise} and Metzen et al. \cite{metzen2017detecting}. The detector appended classifier models of Sterneck et al. \cite{sterneck2021noise} and Metzen et al. \cite{metzen2017detecting} are implemented on the cloud as shown in Fig. \ref{fig:motivation}b. The cloud implementation is performed on a 45nm CMOS Eyeriss DNN accelerator proposed in \cite{chen2016eyeriss} for a fair energy evaluation. As both prior works \cite{sterneck2021noise, metzen2017detecting} perform adversarial detection at the classifier, natural and adversarial data are transmitted from edge to the classifier ($p$= 1 $q$= 1) leading to high $E_{T,A}$. Further, $E_D$ in Sterneck et al. \cite{sterneck2021noise} and Metzen et al. \cite{metzen2017detecting}, are $25\times$ and $64\times$ greater than RobustEdge. This is because Sterneck et al \cite{sterneck2021noise} and Metzen et al \cite{metzen2017detecting} append detector networks at the end of 5th and 11th convolution layers, respectively. Thus, they entail huge computation overhead for detection (4/7 convolution layers and the detector networks). 
\textbf{Comparison under Black-box Attacks} As seen in Fig. \ref{fig:inf_training_energy_pw}b, a 16b QES-trained detector shows high performance against black-box PGD8 attacks compared to Sterneck et al. \cite{sterneck2021noise} and Moitra et al on the CIFAR100 dataset. \cite{moitra2021detectx}. For both prior works, the detector appended to a VGG19 model was trained on natural and adversarial data $x_{adv}$ using the VGG19 model. For RobustEdge, the detector was trained on natural and adversarial data generated using VGG19 model. Additionally, the black-box attacks are generated using a ResNet18 model trained on natural data using SGD.  
\subsection{QES-Training with Limited Training Data}

Fig. \ref{fig:lim_data} shows the AUC scores of detectors created using 1) 10\% 2) 40\% and 3) 60\% of the TinyImagenet dataset. The subsets are created by randomly sampling from the actual dataset. For reference, we also show the performance of the detector trained on the full dataset. In all cases, the detectors are trained and tested on adversarial data created using ResNet18 model (white-box attacks). It can be observed that detectors trained with just 40\% of the training data can achieve AUC scores comparable with the detector trained on the full data. Interestingly, for some attacks (such as GN) merely 10\% of the training data is sufficient for achieving a high performance. Similar observations are made on the CIFAR10 and CIFAR100 datasets.

\subsection{Transferability Across Datasets}

In this section, we explore the following question: \textit{Can a detector trained on dataset A (source dataset) be used to detect adversaries from another dataset B (target dataset)?}.
\noindent\textbf{Methodology: } The source dataset is used to train the weights of a three-layered detector network using QES training. For detector transferability, the layer-specific confidence boundaries ($\mathcal{E}_{i,L}$ and $\mathcal{E}_{i,U}$) and the threshold energies $\mathcal{E}_{n,Th}$ are fine-tuned based on the target dataset (in the ``Confidence Boundary Generation" section of Algorithm \ref{alg:algorithm}) . For this, sample data $s_{nat, target}$ is sampled from the target dataset. $s_{nat, target}$ is forwarded through the detector layers creating $\Psi_{i,s,target}$ energy distributions. $\Psi_{i,s,target}$ is used to compute $\mathcal{E}_{i,L}$, $\mathcal{E}_{i,U}$ and $\mathcal{E}_{n,Th}$. Through our experiments, we find that a size of 200 samples for $s_{nat, target}$ data is sufficient to fine-tune the layer-specific confidence boundary and threshold energy for the target dataset.  


\begin{figure*}[h!] 
\centering
    \begin{tabular}{c}
     \includegraphics[width=0.8\textwidth]{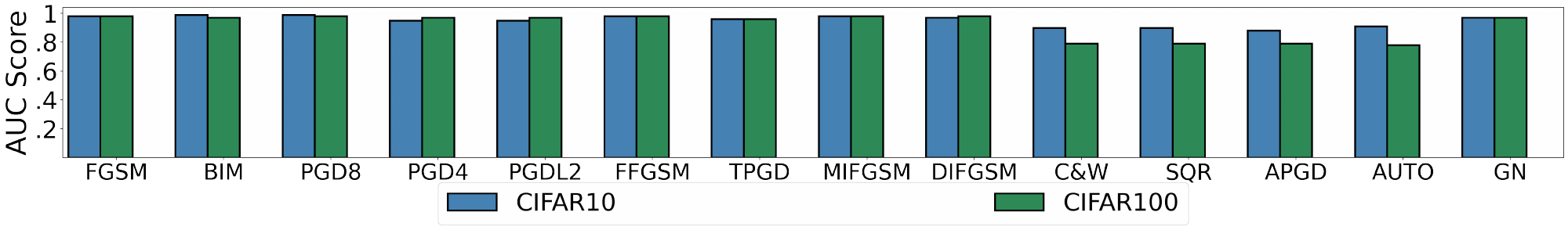}\vspace{-2mm}\\ (a) \\
     
     \includegraphics[width=0.8\textwidth]{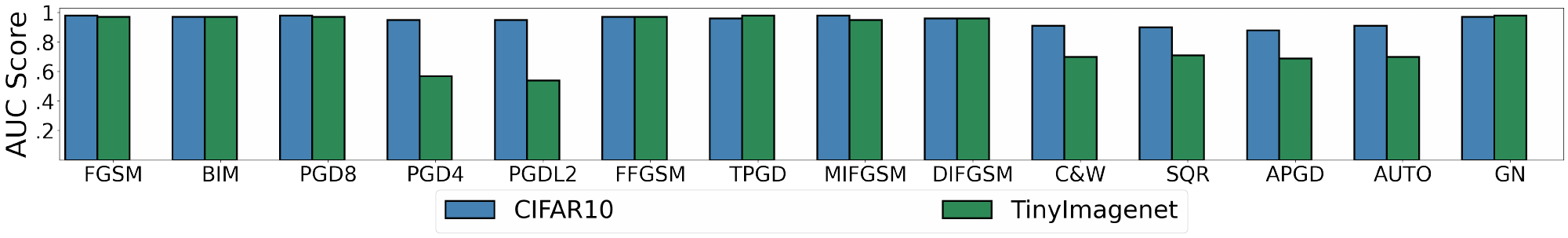}\vspace{-2mm} \\ (b)
     
     \end{tabular}
     \caption{AUC scores for (a) detector trained with source dataset TinyImagenet and transferred to target datasets CIFAR10 and CIFAR100 (b) detector trained with source dataset CIFAR100 and transferred to target datasets CIFAR10 and TinyImagenet}
     \label{transfer_tiny}
\end{figure*}
\noindent\textbf{Transferability Results:} Fig. \ref{transfer_tiny}a shows the detection performance of a detector trained on TinyImagenet as the source dataset and transferred to target datasets CIFAR10 and CIFAR100. Here, all the evaluations are based on white-box attacks. Evidently, the detector shows high AUC scores across different adversarial attacks when transferred to CIFAR10 and CIFAR100 datasets. Interestingly, a detector trained on CIFAR100 as the source dataset transfers well to CIFAR10 dataset but does not transfer well to a TinyImagenet dataset as seen in Fig. \ref{transfer_tiny}b. 


\subsection{Ablation Studies}
\label{detector_arch_sec}
\subsubsection{{\textbf{Detector Architecture and Performance.}}}


\begin{figure}[h!]
    \centering
    \resizebox{\columnwidth}{!}{
    \begin{tabular}{c}
    {\resizebox{0.3\columnwidth}{!}{\begin{tabular}{c|ccc}
    \hline
         Detector & Layer1 & Layer2 & Layer3 \\ \hline
         $\mathcal{D}_1$ & C(3,8) - R & C(8,16) - R & C(16,32)\\
         $\mathcal{D}_2$ & C(3,16) - R & C(16,32) - R & C(32,64) \\ 
         $\mathcal{D}_3$ & C(3,32) - R & C(32,32) - R & C(32,64)\\
          \hline
    \end{tabular}
    }} \\ \includegraphics[width=0.3\columnwidth]{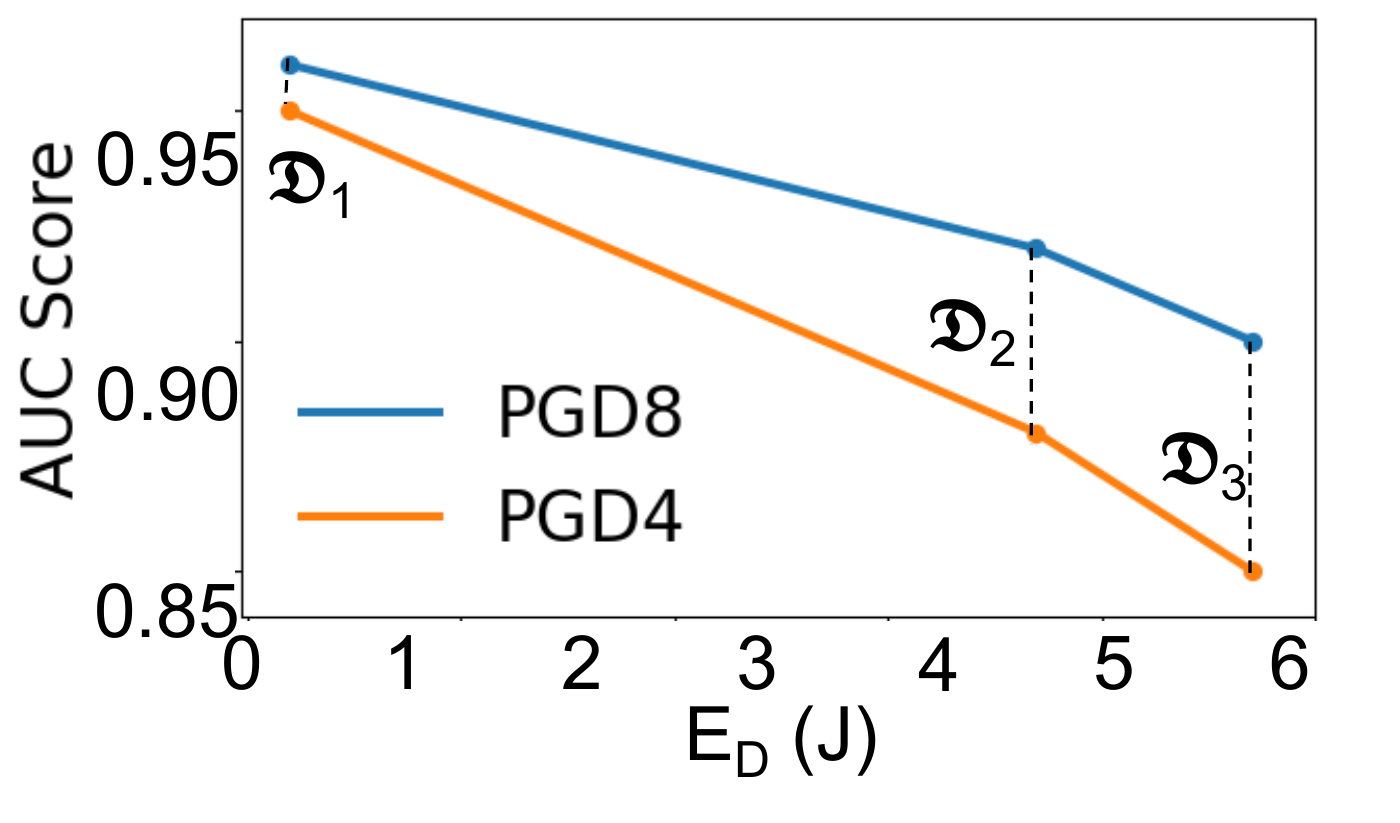}
    
    \end{tabular}}
    
    \caption{(Top) Table showing different 3-layered detector architectures $\mathcal{D}_1$, $\mathcal{D}_2$, $\mathcal{D}_3$. C(n,m)-R denotes a convolution layer with n input and m output channels followed by ReLU layer. (Bottom) AUC scores and $E_D$ for PGD8 and PGD4 attacks for $\mathcal{D}_1$, $\mathcal{D}_2$, $\mathcal{D}_3$ detectors shown in (a). AUC and $E_D$ is computed using $N_{nat}$=$N_{adv}$=10k data sampled from the CIFAR100 training set.}
    \label{fig:detector_arch}
\end{figure}


    
    

In this section we perform an ablation to understand the significance of network architecture on performance and $E_D$. Interestingly, Fig. \ref{fig:detector_arch} shows that a narrow detector $\mathcal{D}_1$ ($\mathcal{D}_1$ has the smallest number of output channels in its network compared to $\mathcal{D}_2$ and $\mathcal{D}_3$) in fact achieves the highest detection compared to wider detectors ($\mathcal{D}_2$, and $\mathcal{D}_3$). Being narrow, $\mathcal{D}_1$ also achieves the lowest $E_D$ compared to other detectors. Further, increasing the detector depth $>$3 only improves the performance marginally while incurring higher $E_D$. Thus, the depth is set to 3. 


\subsubsection{{\textbf{Choosing $K$, $L$ and $U$ values}}}

\begin{figure}[h]
    \centering
        \includegraphics[width=0.9\columnwidth]{ 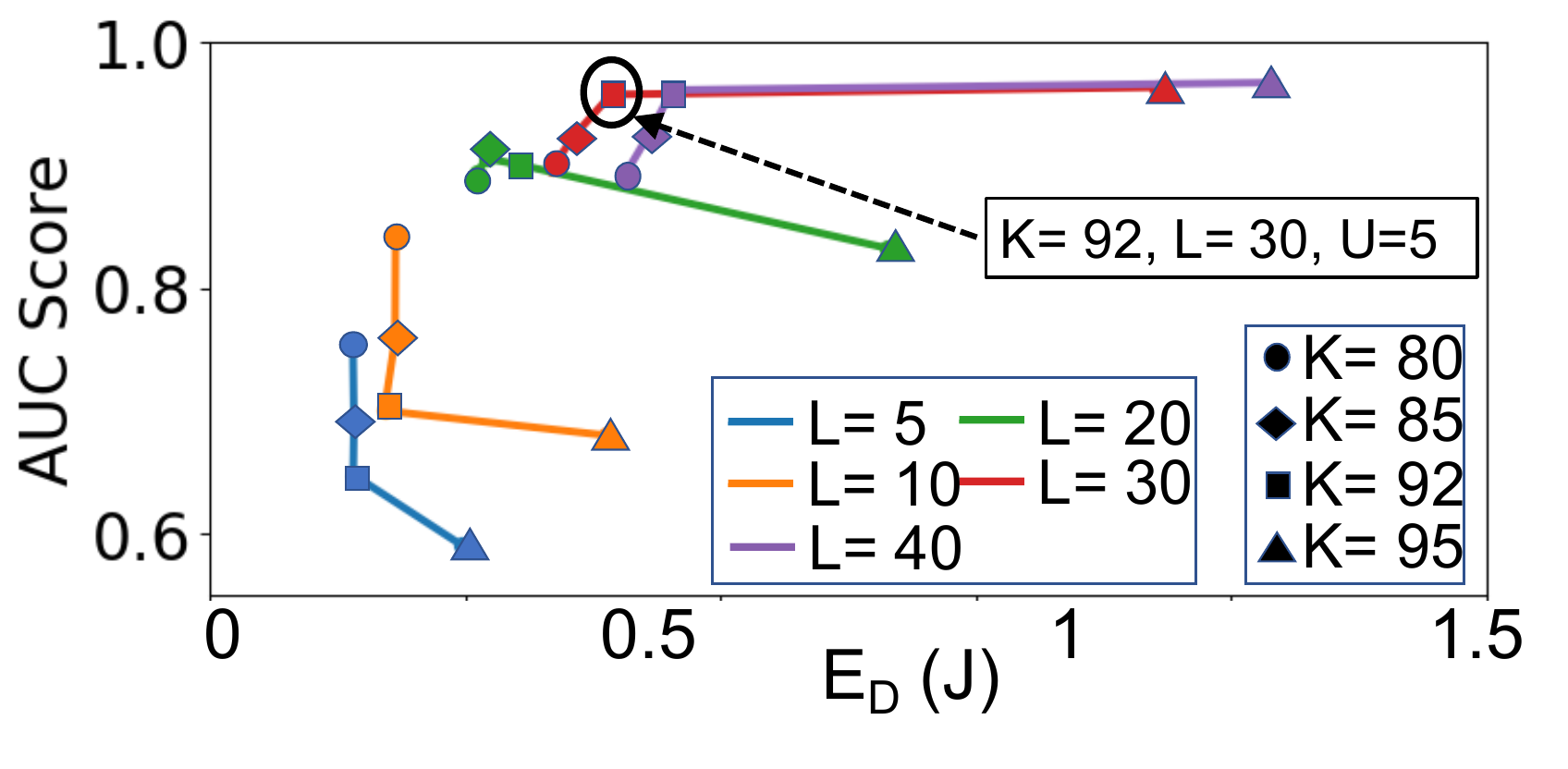} 
          
     
    \caption{Figure showing the AUC scores and $E_D$ of the 16b Detector Z against different values of $L$ and $K$ at $U$= 5. All results correspond to PGD4 attacks on data from the CIFAR100 training set. Here,  $N_{nat}$=$N_{adv}$=10k. }
    \label{fig:klu_params} 
\end{figure}

\begin{figure}[h]
    \centering
        \includegraphics[width=0.4\columnwidth]{ 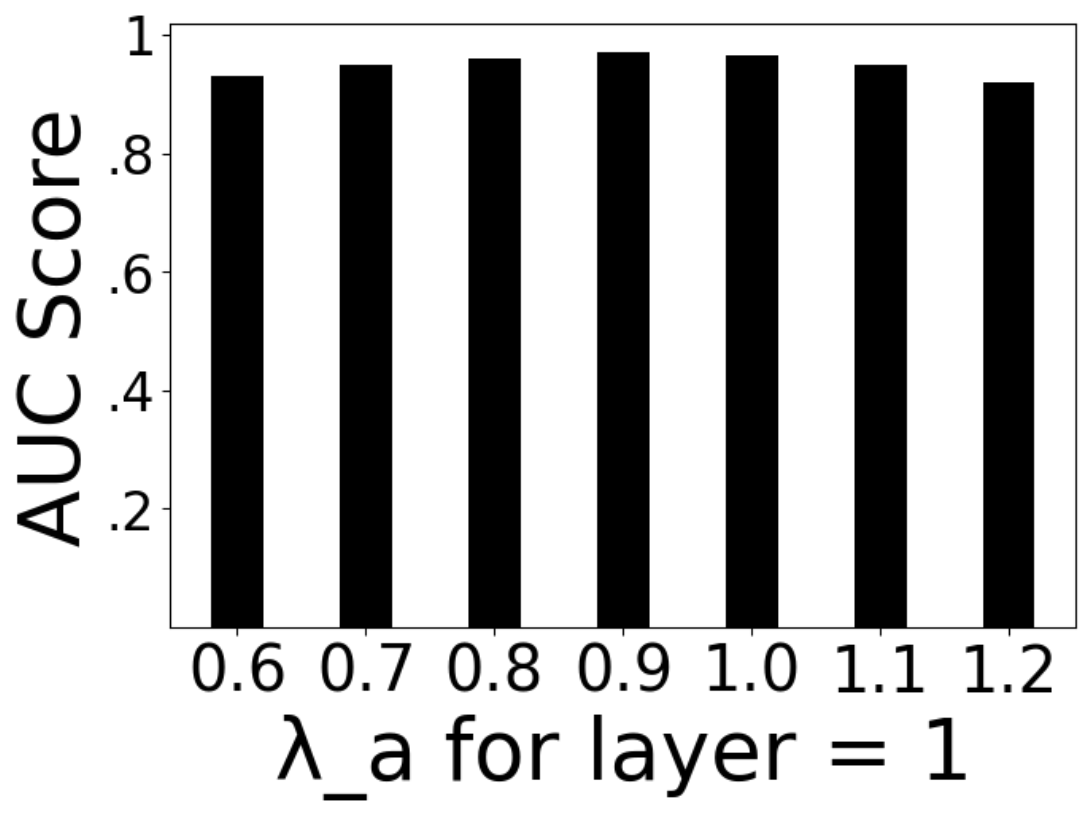} 
          
     
    \caption{AUC scores for different 16b detectors trained with different $\lambda_a$ values at $\lambda_n$= 0.1. AUC scores correspond to PGD4 attacks (CIFAR100 training set data) for Detector Z and $K$=92, $L$= 30, $U$= 5.}
    \label{fig:lambda_ablation} 
\end{figure}

The values of $K$, $L$ and $U$ determine the performance and energy efficiency of the detector (see Section \ref{early_exit} for definition). As seen in Fig. \ref{fig:klu_params}, we achieve the best tradeoff between $E_D$ and adversarial detection for $K$=92, $L$= 30 and $U$= 5. Smaller $L$ ($L$= 5, 10, 20) implies confident early exits in the initial layers leading to low $E_D$ but poor AUC scores irrespective of the $K$ value. At larger $L$ ($L$= 30 or 40), increasing $K$ leads to slightly higher AUC scores at the cost of higher $E_D$.
\subsubsection{\textbf{Choosing $\lambda_n$ and $\lambda_a$}} 

Fig. \ref{fig:lambda_ablation} shows the AUC scores for a 16b detector trained with different $\lambda_a$ values for layer 1 (refer Section \ref{early_exit} for definitions of $\lambda_n$ and $\lambda_a$). Highest AUC score is seen for $\lambda_a = 0.9 $. For layers 2 and 3, the $\lambda_a$ values must be higher than layer 1 to achieve higher \textit{energy} separation. Therefore, $\lambda_a$ for layer 2 and 3 are incremented by 0.4 and 1.1 with respect to layer 1's $\lambda_a$. The best AUC score is obtained for $\lambda_n$= 0.1 and $\lambda_a$=0.9, 1.3 and 2 for training layers 1, 2 and 3, respectively. 

\subsection{Raspberry Pi 4 Implementation}
\begin{table}[h!]
    \centering
    \begin{tabular}{|l|c|} \hline
        Compute Current & 120mA \\ \hline
        Latency L1, (L1+L2), (L1+L2+L3) & 0.3ms, 0.6ms, 1.3ms \\ \hline
        WiFi Current  & 160mA \\ \hline
        WiFi Latency (CIFAR100) & 3.5ms \\ \hline
        Supply Voltage & 12V \\ \hline
        \multicolumn{2}{|c|}{Energy = Voltage $\times$ Current $\times$ latency} \\\hline
        
          
    \end{tabular}
    \caption{Current, voltage and latency of the Raspberry-Pi implementation. Latency with early exits at layer1, layer2 and layer3 are shown by L1, (L1+L2) and (L1+L2+L3), respectively.}
    \label{tab:raspi_char}
\end{table}
\begin{figure}[h!]
    \centering
    \includegraphics[width=0.4\linewidth]{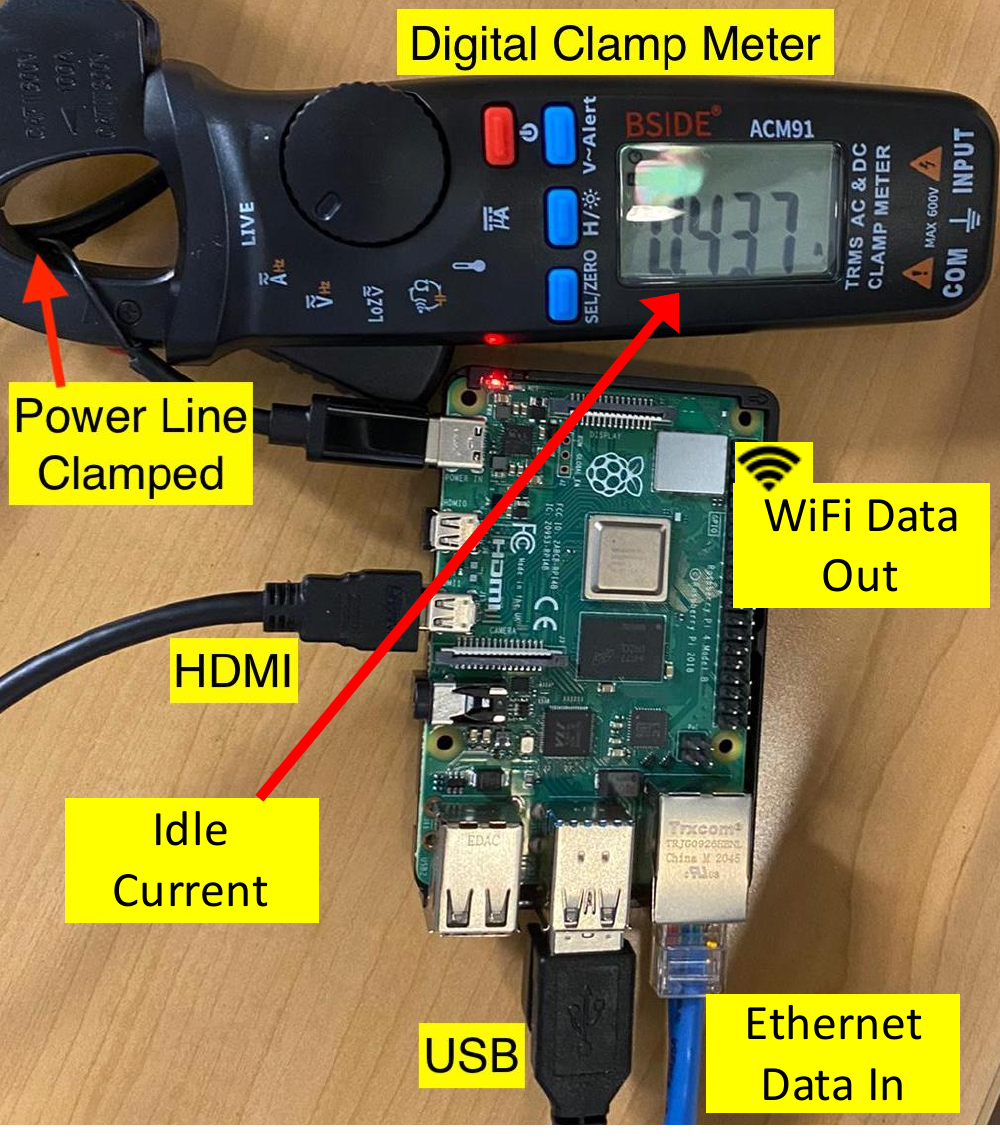}
    \caption{Figure showing the Raspberry-Pi setup used for the demonstration.}
    \label{fig:setup}
\end{figure}
\begin{figure}[h!]
    \centering
    \begin{tabular}{c}
         \includegraphics[width=0.6\columnwidth]{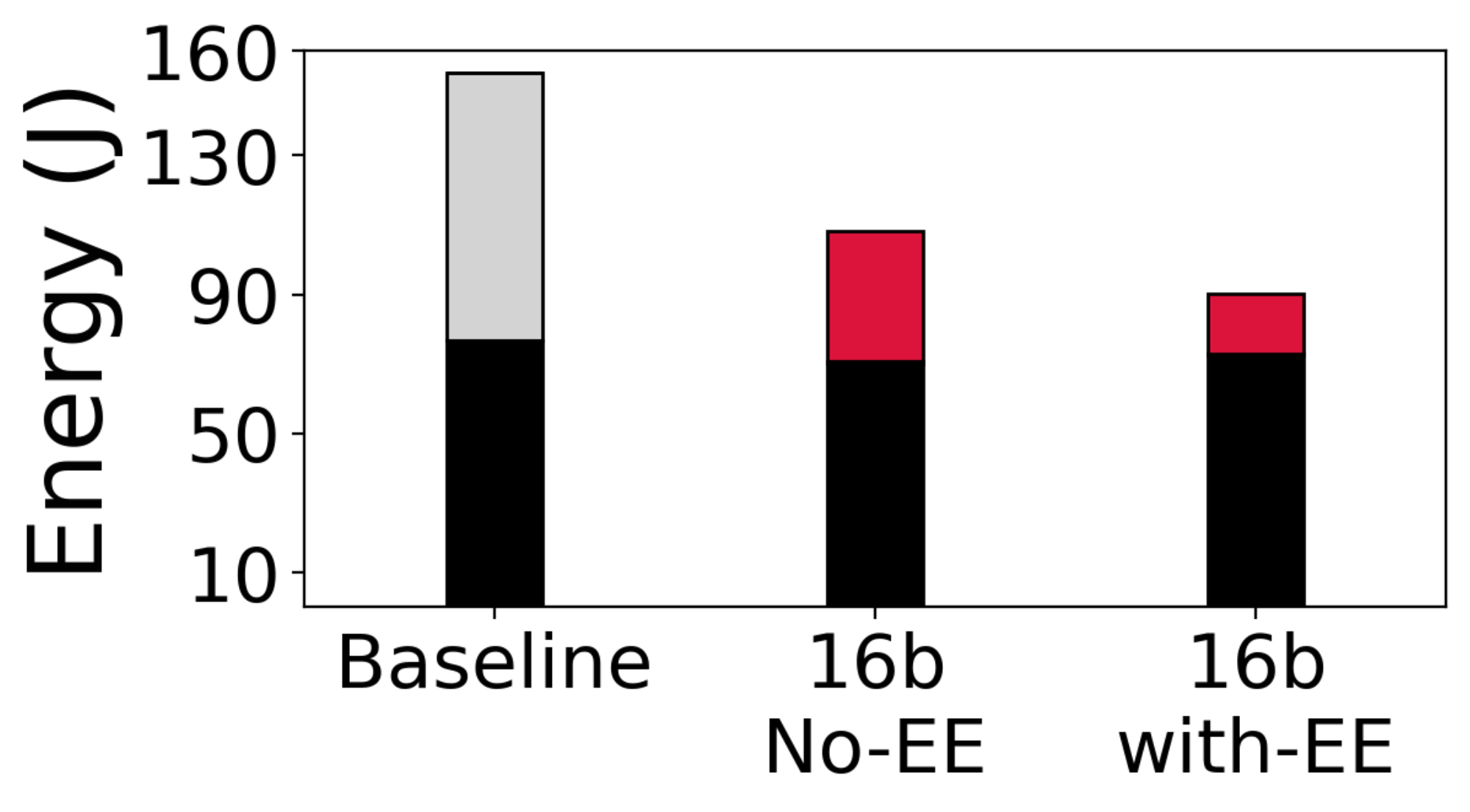} \vspace{-2mm} \\
         \hspace{7mm}\includegraphics[width=0.4\columnwidth]{figure/PGD_legend.pdf}  
    \end{tabular}
    \caption{$E_{T,A}$, $E_{T,N}$ and $E_D$ comparison between ``Baseline" and classifier-edge system with a 16b QES-trained detector implemented on a Raspberry-Pi-4 edge device. The results correspond to the CIFAR100 dataset under PGD4 attacks. The energy is computed for processing 10k natural and 10k adversarial samples. }
    \label{fig:raspi_pgd4}
\end{figure}

We implement the 16b 3-layered QES-trained detector on the Raspberry-Pi 4 platform. The detectors are deployed using Tensorflow v2.5 running on Raspberry Pi OS. Table \ref{tab:raspi_char} shows the current, voltage and latency of the implementation. The data transmission (to a remote cloud GPU (RTX2080Ti) in our case) takes place over WiFi. Fig. \ref{fig:setup} shows the Raspberry-Pi 4 setup. The HDMI and USBs are used for the display and keyboard/mouse, respectively. The ethernet is used for adversarial and natural data acquisition. The adversarial detection occurs inside Raspberry-Pi 4's Cortex A7 processor and the natural data  transmission to the cloud occurs via the in-built WiFi module. We use a current clamp meter to measure the current drawn by the Raspberry-Pi. In Fig. \ref{fig:setup} the clamp meter shows the idle-state current drawn by the Raspberry-pi-4. 
During adversarial detection, the current rises to 557mA ($\therefore$ compute current = 120mA). While the current rises to 720mA when data transmission over WiFi is activated along with adversarial detection ($\therefore$ WiFi current = 160mA). Additionally to compute the latency, we measure the run time of detection using the $time()$ function in Tensorflow. The latency values are shown in Table \ref{tab:raspi_char}. For a detector without early exit, the latency equals (L1+L2+L3) as shown in Table \ref{tab:raspi_char}.

From Fig. \ref{fig:raspi_pgd4}, we find that the QES-trained detector with ``early detection and exit" entails negligible cost due to adversarial data transmission ($E_{T,A}$= 0). However, the detection cost ($E_D$) is significantly high. This is because unlike systolic array accelerator, Raspberry-Pi's architecture entails higher number of energy-intensive DRAM accesses. It must be noted that ``early-detection and exit" is highly advantageous here as it reduces the detection cost by 66\% compared to detection without early-exit.


\section{Conclusion}
RobustEdge proposes QES-training with ``Early detection and Exit" to perform low power edge-based adversarial detection. The QES-trained detector is implemented on a 45nm CMOS digital hardware accelerator. The edge detector achieves state-of-the-art AUC score $>$ 0.9 against standard gradient-based white-box and black-box attacks at extremely small detection energy ($<25\times$ compared to prior works). Interestingly, a detector trained on gradient-based attacks has high detection score against score-based attacks (AUC score $>$0.7). Further, it is shown to improve the energy efficiency of a realistic cloud-edge system by $>166\times$ compared to prior works. Being transferable across datasets, a detector trained on one dataset can be reused to detect adversaries on another dataset thereby saving the training cost and adding to the energy efficiency. 


\section*{Acknowledgement}

This work was supported in part by CoCoSys, a JUMP2.0 center sponsored by DARPA and SRC, Google Research Scholar Award, the National Science Foundation CAREER Award, TII (Abu Dhabi), and the DoE MMICC center SEA-CROGS (Award\#DE-SC0023198).

\bibliographystyle{IEEEtran}
\bibliography{egbib.bib}


\end{document}